\pdfoutput=1

\documentclass[aps,twocolumn,preprintnumbers,amssymb,nobibnotes,nofootinbib,longbibliography,superscriptaddress,10pt]{revtex4}

\usepackage[utf8]{inputenc}

\usepackage{titlesec}
\usepackage{bbold}

\usepackage{graphicx,epsf}
\usepackage{amsmath,amsfonts,amssymb,amsbsy,mathrsfs}

\usepackage{caption}
\usepackage{subcaption}
\usepackage{physics}

\usepackage{color,xcolor}

\usepackage{array,multirow}
\usepackage{stackrel}
\usepackage{tikz}
\usetikzlibrary{calc}

\usepackage{ifdraft}
\usepackage[obeyFinal]{easy-todo}

\usepackage{placeins}
\usepackage{slashed,shuffle} 
\usepackage{adjustbox}

\usepackage{comment}
\usepackage[normalem]{ulem}

\usepackage{slashed}
\usepackage{hyperref}
\hypersetup{
    colorlinks=true,
    allcolors={blue!60!black}
}

\usepackage{cleveref}
\usepackage{pifont}
\newcommand{\cmark}{\ding{51}}%
\newcommand{\xmark}{\ding{55}}%

\newcolumntype{C}[1]{>{\centering\arraybackslash}m{#1}}

\newcommand{\eps}{\epsilon}
\def\trFive{{\rm tr}_5}

\def\offShellScale{p_1^2}

\newcommand{\ii}{\mathrm{i}}

\newcommand*{\nsqrt}{\Sigma_5}

\newcommand{\fsigma}{\mathcal{F}_{\vb*{\Sigma_5}}}
\newcommand{\fnosigma}{\bar{\mathcal{F}}_{\vb*{\Sigma_5}}}

\titleformat{\section}{\centering\normalsize\normalfont\bf}{\thesection}{1em}{}
\makeatletter\DeclareRobustCommand*{\bfseries}{\not@math@alphabet\bfseries\mathbf\fontseries\bfdefault\selectfont\boldmath}\makeatother

\begin{document}

\title{All Two-Loop Feynman Integrals for Five-Point One-Mass
Scattering}
\preprint{CERN-TH-2023-119, ZU-TH 26/23, LAPTH-036/23}


%
\author{Samuel~Abreu}
\affiliation{CERN, Theoretical Physics Department, CH-1211 Geneva 23, Switzerland}
\affiliation{Higgs Centre for Theoretical Physics, School of Physics and Astronomy,
The University of Edinburgh, Edinburgh EH9 3FD, Scotland, UK}
\author{Dmitry~Chicherin}
\affiliation{LAPTh, Universit\'e Savoie Mont Blanc, CNRS, B.P.~110, F-74941 Annecy-le-Vieux, France}
\author{Harald~Ita}
\affiliation{Paul Scherrer Institut, Forschungsstrasse 111, 5232 Villigen, Switzerland}
\affiliation{ICS, University of Zurich, Winterthurerstrasse 190, Zurich, Switzerland}
\author{Ben~Page}
\affiliation{CERN, Theoretical Physics Department, CH-1211 Geneva 23, Switzerland}
\author{Vasily~Sotnikov}
\affiliation{Physik-Institut, University of Zurich, Winterthurerstrasse 190, 8057 Zurich, Switzerland}
\affiliation{Department of Physics and Astronomy, Michigan State University, 567 Wilson Road, East Lansing, MI 48824, USA}
\author{Wladimir~Tschernow}
\affiliation{Physikalisches Institut, Albert-Ludwigs-Universit\"at Freiburg,
D-79104 Freiburg, Germany}
\author{Simone~Zoia}
\affiliation{Dipartimento di Fisica and Arnold-Regge Center, Universit\`a di Torino, 
and INFN, Sezione di Torino, Via P.~Giuria 1, I-10125 Torino, Italy}

\begin{abstract}
We compute the complete set of two-loop master integrals for the scattering of
four massless particles and a massive one.
Our results are ready for phenomenological applications, 
removing a major obstacle to the computation of complete next-to-next-to-leading order (NNLO) QCD
corrections to processes 
such as the production of a $H/Z/W$ boson in association with two jets at the LHC.
Furthermore, they open the door to new investigations into the structure of
quantum-field theories and provide precious analytic data for
studying the mathematical properties of Feynman integrals.
\end{abstract}

\maketitle

Feynman integrals play a central role in obtaining precise predictions in
quantum-field theory (QFT). They are also of great
mathematical interest, giving rise to noteworthy classes of special functions.
Advances in the calculation of Feynman integrals have led to
new insights into the mathematical properties
of these functions, as well as to new results in formal studies of QFTs
and (beyond) Standard-Model phenomenology. 
While the calculation of two-loop five-point Feynman integrals is an active area of 
research~\cite{Gehrmann:2015bfy,Papadopoulos:2015jft,Abreu:2018rcw,
Abreu:2018aqd,Chicherin:2018old,Abreu:2020jxa,Chicherin:2020oor,
Canko:2020ylt,Abreu:2021smk,Kardos:2022tpo,Badger:2022hno,Hidding:2022ycg},
a complete set of integrals is only available for massless 
particles~\cite{Gehrmann:2015bfy,Papadopoulos:2015jft,Abreu:2018rcw,
Abreu:2018aqd,Chicherin:2018old,Abreu:2020jxa,Chicherin:2020oor}.
In this letter, we advance
the state of the art by computing
all Feynman integrals necessary to describe the scattering of four massless
particles and a massive one at two loops.

We obtain a representation for the Feynman integrals in dimensional regularization 
that exhibits their analytic
structure and allows for a stable and efficient numerical evaluation. This
is achieved by finding bases of pure integrals~\cite{Arkani-Hamed:2010pyv} that
satisfy differential equations (DEs)~\cite{Kotikov:1990kg,Kotikov:1991pm,Bern:1993kr,Remiddi:1997ny,Gehrmann:1999as}
in canonical form~\cite{Henn:2013pwa}, explicitly displaying all 
singularities of the integrals.
Despite recent progress~\cite{Chicherin:2018old,Dlapa:2020cwj,Henn:2020lye,
Dlapa:2022wdu,Gorges:2023zgv,Dlapa:2021qsl,Chen:2022lzr}, finding a pure basis is still a major
bottleneck. We followed the approach
of refs.~\cite{Abreu:2018rcw,Abreu:2020jxa}, building on modular 
arithmetic to bypass intermediate computational bottlenecks~\cite{vonManteuffel:2014ixa,Peraro:2016wsq}.
Using Chen iterated integrals \cite{Chen:1977oja},
we solve the DEs at each order in the dimensional regulator
in terms of a minimal set of functions, called \emph{(one-mass) pentagon functions}.
We demonstrate that this solution is efficient and numerically stable over phase
space and therefore ready for phenomenological application.
A \texttt{C++} library \cite{PentagonFunctions:cpp} for the
evaluation of the pentagon functions makes
our results accessible to the whole community.
Previous approaches for constructing such solutions~\cite{Gehrmann:2018yef,Chicherin:2020oor,Badger:2021nhg,Chicherin:2021dyp}
relied on the possibility of representing them through multiple polylogarithms (MPLs)~\cite{Goncharov:2010jf}, but
it is generally unclear if such a representation exists \cite{Duhr:2020gdd}.
We show that this is not required, and
only one evaluation of the integrals at a precision comparable to that
at which we evaluate the pentagon functions is sufficient. 
Non-planar integrals introduce added complexity:
some exhibit non-analytic behavior or a logarithmic singularity within 
the physical scattering region. We isolate this behavior in the pentagon functions,
and extend the numerical methods of refs.~\cite{Chicherin:2020oor,Chicherin:2021dyp} to deal with it.

Our results open the door to new explorations in many different directions.
On the analytic side, this is the first complete set of two-loop integrals allowing us to explore
the (extended) Steinmann relations \cite{Steinmann:thesis,Steinmann:1960,Cahill:1973qp,Caron-Huot:2016owq,
  Dixon:2016nkn,Caron-Huot:2018dsv,Hannesdottir:2022xki}.
Moreover, these integrals make it possible to study how unexplained observations
of analytic cancellations in gauge-theory amplitudes~\cite{Badger:2021nhg,
Badger:2021ega, Abreu:2021asb, Badger:2022ncb} extend beyond the leading-color approximation.
In formal studies of QFTs, they
have already been central in bootstrapping results in $\mathcal{N}=4$ super Yang-Mills theory (sYM), leading to new 
conjectures \cite{Dixon:2021tdw,Dixon:2022xqh}. 
Finally, all two-loop Feynman integrals for the production of a massive
boson in association with two jets at hadron colliders are now readily available, 
removing a main bottleneck for these important processes in LHC physics for which
more precise theoretical predictions are urgently needed. Indeed, it is now possible to
go beyond the leading-color approximation for $W$ + 2-jet processes
\cite{Badger:2021nhg,Abreu:2021asb,Hartanto:2022qhh,Hartanto:2022ypo,Buonocore:2022pqq},
and obtain two-loop results for crucial processes such as $H$ + 2 jets which 
are intrinsically non-planar.

\section{Notation and Conventions}\label{sec:notation}
\begin{figure}
	\centering
	\begin{subfigure}{0.15\textwidth}\centering
		\includegraphics[scale=0.25]{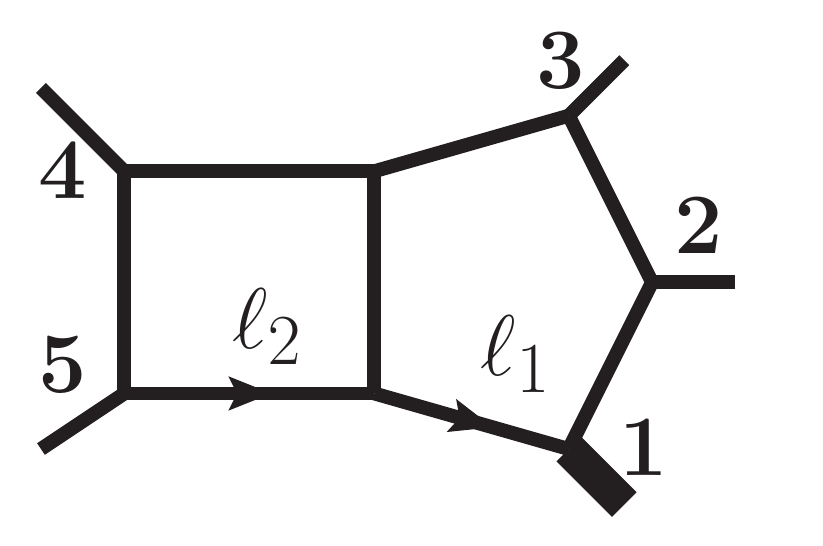}
		\caption{PBmzz} 
		\label{fig:pbmzz}
	\end{subfigure}
	\begin{subfigure}{0.15\textwidth}\centering
		\includegraphics[scale=0.25]{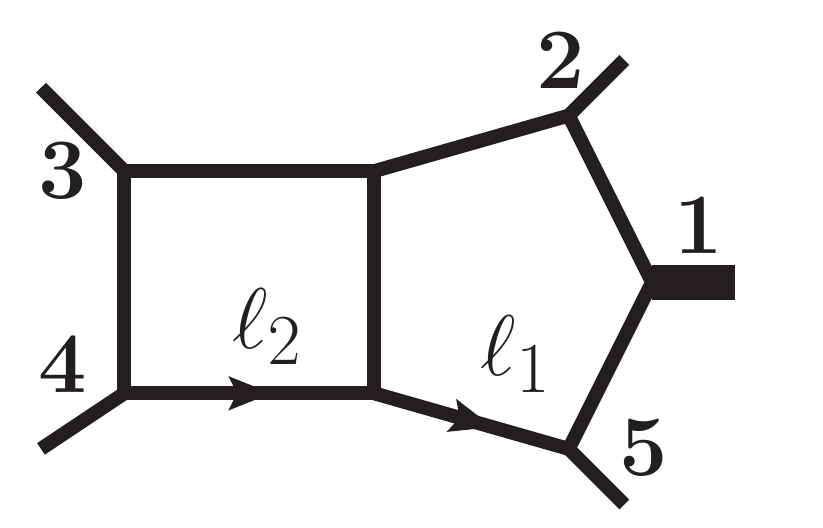}
		\caption{PBzmz}
		\label{fig:pbzmz}
	\end{subfigure}
	\begin{subfigure}{0.15\textwidth}\centering
		\includegraphics[scale=0.25]{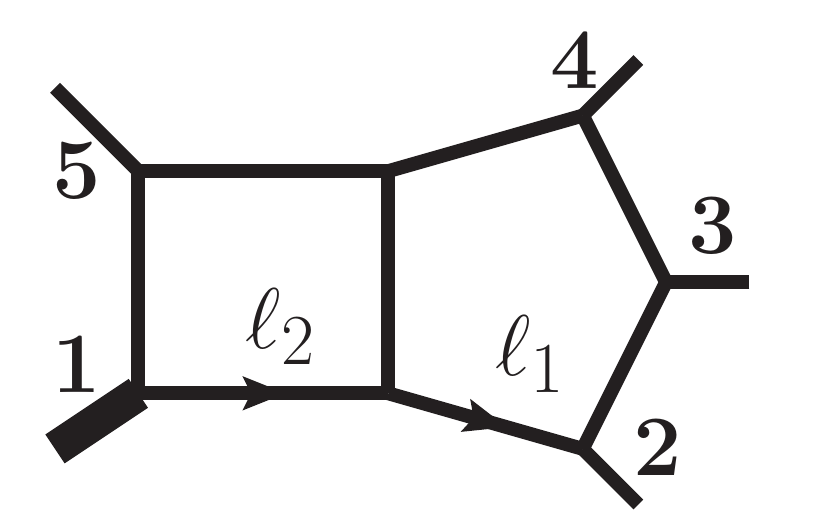}
		\caption{PBzzz}
		\label{fig:pbzzz}
	\end{subfigure}
	\begin{subfigure}{0.15\textwidth}\centering
		\includegraphics[scale=0.25]{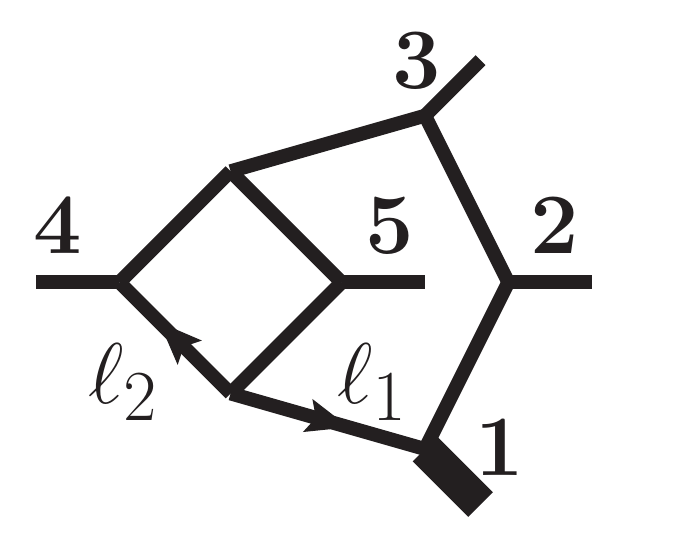}
		\caption{HBmzz} 
		\label{fig:hbmzz}
	\end{subfigure}
	\begin{subfigure}{0.15\textwidth}\centering
		\includegraphics[scale=0.25]{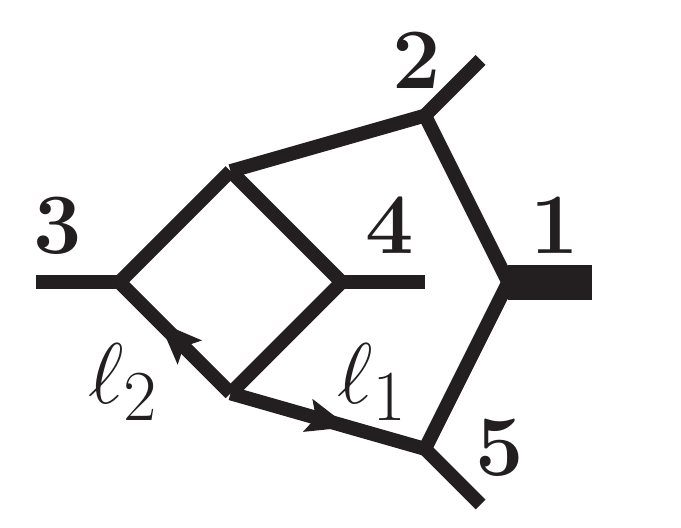}
		\caption{HBzmz}
		\label{fig:hbzmz}
	\end{subfigure}
	\begin{subfigure}{0.15\textwidth}\centering
		\includegraphics[scale=0.25]{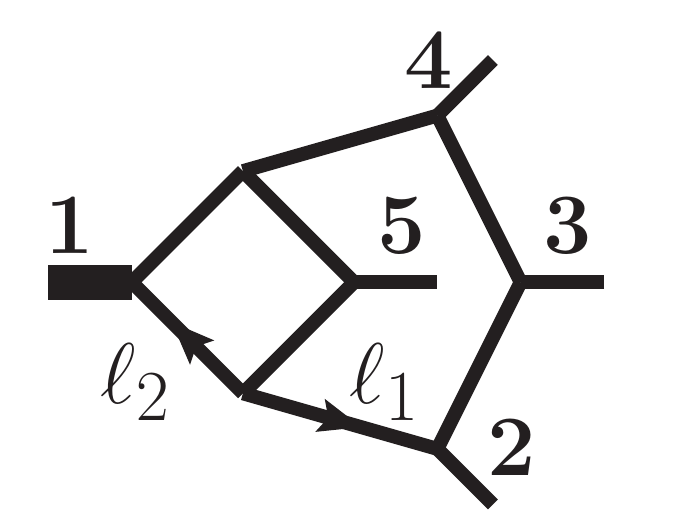}
		\caption{HBzzz}
		\label{fig:hbzzz}
	\end{subfigure}
	\begin{subfigure}{0.20\textwidth}\centering
		\includegraphics[scale=0.25]{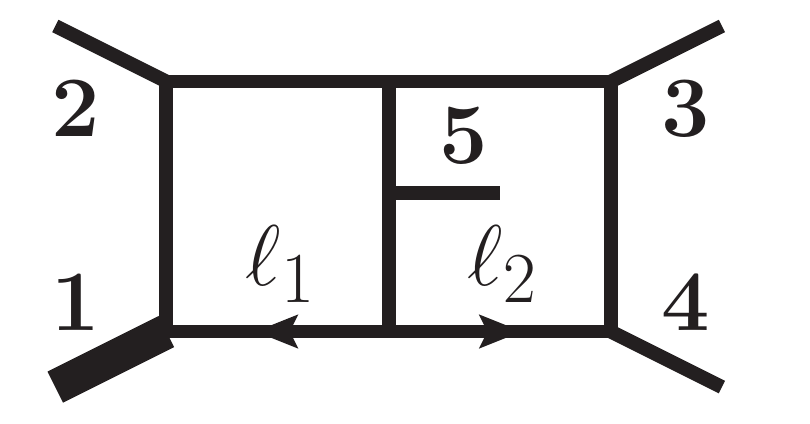}
		\caption{DPmz} 
		\label{fig:mz}
	\end{subfigure}
	\begin{subfigure}{0.20\textwidth}\centering
		\includegraphics[scale=0.25]{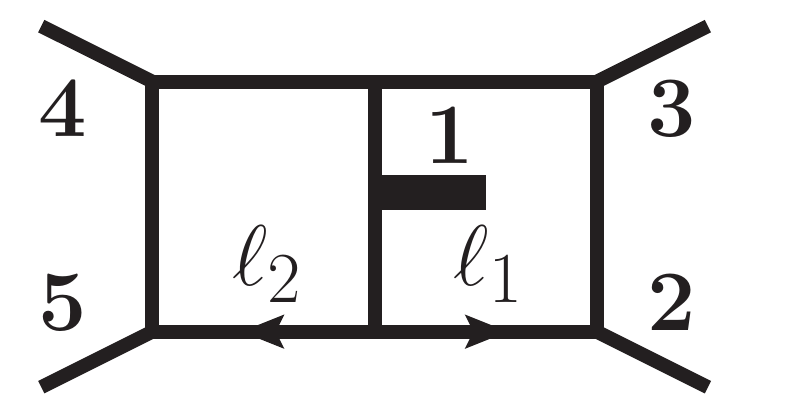}
		\caption{DPzz}
		\label{fig:zz}
	\end{subfigure}
	\caption{Two-loop five-point one-mass families. 
	The thick external line denotes the massive external leg.} 
	\label{fig_families_int}
\end{figure}

%
We define the momenta of external particles as $p_i$, $i=1,\ldots,5$,  where 
$p_1^2 \neq 0$ and $p_{i}^2=0$ for $i=2,\ldots,5$.  
For a fixed ordering of the massless legs, there are three planar penta-box (PB)
families, three non-planar hexa-box (HB) families
and two non-planar double-pentagon (DP) families
that we depict in \cref{fig_families_int}, as well as a factorizable planar
topology.
The factorizable, PB and HB families have already been
studied in the literature \cite{Papadopoulos:2015jft,Abreu:2020jxa,
Canko:2020ylt,Abreu:2021smk,Kardos:2022tpo}. 
Here we define the DP families.
Integrals in these families can generically be written as
\begin{equation}\label{eq:intGen}
I [\vec \nu] =  e^{2 \epsilon \gamma_{\rm E}}
\int \frac{{\rm d}^D\ell_1}{\mathrm{i} \pi^{D/2}}\frac{{\rm d}^D\ell_2 }{\mathrm{i} \pi^{D/2}} \;
\frac{\rho_{9}^{-\nu_9}\,\,\rho_{10}^{-\nu_{10}}\,\,\rho_{11}^{-\nu_{11}}}
{\rho_{1}^{\nu_1} \;  \cdots \;
\,\rho_{8}^{\nu_8}}\,,
\end{equation}
where we set $D=4-2\epsilon$, and $\vec \nu$ is a vector of integers
with the restriction that $\nu_9,\nu_{10},\nu_{11}\leq0$. Explicit
expressions for the $\rho_i$ 
are given in ancillary files~\cite{abreu_samuel_2023_8082812}.

There are six independent variables
$s_{ij}=(p_i+p_j)^2$, which we choose to be
\begin{equation} \label{eqn:orderedInvariants}
\vec s= \{ \offShellScale{} \,,
s_{12}\,, s_{23}\,,s_{34}\,, s_{45}\,, s_{15}\}\,.
\end{equation}
Together with the parity-odd object
\begin{equation}
\label{eq:tr5}
\trFive{} = 4 \mathrm{i} \varepsilon_{\alpha\beta\gamma\delta}
\,p_1^\alpha p_2^\beta p_3^\gamma p_4^\delta\,,
\end{equation}
they fully specify a point in the five-particle phase space.
Singularities of Feynman integrals are
located at zeroes of certain determinants, see e.g.~refs.~\cite{Abreu:2017ptx,
Arkani-Hamed:2017ahv,Hannesdottir:2021kpd,Hannesdottir:2022bmo,
Bourjaily:2022vti}. 
Three cases play a special role here: the three
and five-point Gram determinants
\begin{align}\begin{split}
	\label{eq:gram35}
	\Delta_3=&\,-\det G(p_1, p_2+p_3) \,, \\
	\Delta_5 =&\,  \det G(p_1,p_2,p_3,p_4) \,,
\end{split}\end{align}
where $G(q_1,\ldots,q_n)=2\{q_i\cdot q_j\}_{i,j\in\{1,\ldots,n\}}$,
and the polynomial \cite{Abreu:2021smk}
\begin{align}\begin{split}\label{eq:newRoot}
	\nsqrt=&\,(s_{12} s_{15} - s_{12} s_{23}  
	- s_{15} s_{45} + s_{34} s_{45} + s_{23} s_{34})^2\\	
	&- 4 s_{23}  s_{34} s_{45} (s_{34} - s_{12} - s_{15})\,.
\end{split}\end{align}
While $\Delta_5 = \trFive^2$,
relating $\trFive{}$ to $\sqrt{\Delta}_5$ precisely is a subtle issue.
We adopt the convention of ref.~\cite{Abreu:2021smk}
to only use $\sqrt{\Delta_5}$ in the pure integrals' definitions.

Fig.~\ref{fig_families_int} shows a fixed ordering of the massless
legs, but we consider the 
set of integrals closed under all permutations of these legs.
While $\Delta_5$ is invariant under these permutations, 
there are three different permutations of $\Delta_3$, denoted
$\Delta_3^{(k)}$, and six different permutations of $\nsqrt$, denoted
by $\nsqrt^{(k)}$. Expressions for the 
$\Delta_3^{(k)}$, $\nsqrt^{(k)}$ and $\Delta_5$ are given
in ancillary files~\cite{abreu_samuel_2023_8082812}.

\section{Analytic Differential Equations}\label{sec:diffEq}
%

We follow refs.~\cite{Abreu:2018aqd,Abreu:2018rcw,
Abreu:2020jxa,Abreu:2021smk}, where analytic DEs~\cite{Kotikov:1990kg,Kotikov:1991pm,Bern:1993kr,
Remiddi:1997ny,Gehrmann:1999as} in canonical form~\cite{Henn:2013pwa} 
are obtained from numerical samples.  
We focus on the DPmz and DPzz families, for 
which we obtain canonical DEs for the first time. Any integral in the DPmz 
(DPzz) family can be written as a linear combination of 142 (179) master 
integrals. The top sectors, with 8 propagators and 9 
master integrals each, were previously unknown.  
All integration-by-parts (IBP) reductions~\cite{Tkachov:1981wb,
Chetyrkin:1981qh,Laporta:2000dsw} are performed
within \texttt{FiniteFlow} \cite{Peraro:2019svx} (interfaced to
\texttt{LiteRed}~\cite{Lee:2012cn,Lee:2013mka}), and checked with
\texttt{Kira 2.0} \cite{Klappert:2020nbg} and \texttt{FIRE6}
\cite{Smirnov:2019qkx}.

Let $\vec{g}_\tau$ denote a vector whose entries form a 
pure~\cite{Arkani-Hamed:2010pyv} basis of master integrals 
for a family of integrals $\tau$.
It satisfies a DE in canonical form~\cite{Henn:2013pwa}
\begin{equation} \label{eq:DE_can}
  \dd{\vec{g}_\tau}=\epsilon \, \mathbf{M} \cdot \vec{g}_\tau\,,\qquad \mathbf{M}=\sum_i \mathbf{M}_i\, \dd{\log W_i}\,,
\end{equation} 
where the $W_i$ are the letters of the (symbol) 
alphabet~\cite{Goncharov:2010jf} associated with $\vec{g}_\tau$. While the $W_i$ are algebraic functions of
$\vec s$, the matrices ${\bf M}_i$ are matrices of rational numbers.
Finding a pure basis is still the most challenging aspect in
obtaining DEs in canonical form. 
We construct educated guesses for pure bases following the ideas of
refs.~\cite{Abreu:2018aqd,Chicherin:2018old,Abreu:2020jxa,Abreu:2021smk},
and test candidate bases by evaluating their derivatives at numerical 
points and verifying the factorization of~$\epsilon$. 
Once a pure basis is found, we proceed as in
section~4 of ref.~\cite{Abreu:2020jxa} to determine
that the alphabet for the DPmz and DPzz families is 
contained within the one obtained in ref.~\cite{Abreu:2021smk}.
DPmz and DPzz have 62 and 74 letters respectively.
As in ref.~\cite{Abreu:2020jxa}, we fit the matrices ${\bf M}_i$ 
from numerical evaluations on a finite field. 
Our results for the pure bases, the alphabet (closed under all 
permutations of the massless legs), and the analytic DEs can be found in ancillary 
files~\cite{abreu_samuel_2023_8082812}.
Some pure integrals were simplified
with ideas from ref.~\cite{Bourjaily:2021hcp}.

\section{Construction of One-Mass Pentagon Functions} 
\label{sec:pf_construction} 

The \emph{(one-mass) pentagon
functions} are a basis of special functions to express all one- and two-loop
five-point integrals with an external massive leg, up to the order 
in~$\epsilon$ required to compute two-loop corrections.
The one-loop and planar two-loop integrals have been previously 
considered in ref.~\cite{Chicherin:2021dyp}.
Here we discuss a novel approach, suitable for the non-planar families, that
overcomes bottlenecks of previous strategies.

We start by considering the $\epsilon$ expansion of the master
integrals for each family $\tau$ in \cref{fig_families_int}. To cover all integrals
relevant for an amplitude, we consider all $4!$ permutations $\sigma$ of the
massless momenta, denoting the corresponding master integrals by $\vec
g_{\tau,\sigma}$. 
We normalize the $\vec g_{\tau,\sigma}$ so that they 
have the expansion
\begin{align} \label{eq:g_exp}
	\vec g_{\tau,\sigma}(\vec{s}) = \sum_{w \geq 0} \eps^{w} 
	\vec g^{(w)}_{\tau,\sigma}(\vec{s}) \,.
\end{align}
We obtain the DEs for all $\vec g_{\tau,\sigma}$ by permuting those in the
standard ordering, and use them to write the $\vec
g^{(w)}_{\tau,\sigma}(\vec{s})$ as $\mathbb{Q}-$linear combinations of Chen
iterated integrals~\cite{Chen:1977oja},
\begin{align} \label{eq:iterated_integral}
\!\!\![W_{i_1},\ldots,W_{i_w}]_{\vec{s}_0}(\vec{s})= \!\int_\gamma [W_{i_1},\ldots,W_{i_{w-1}}]_{\vec{s}_0} \mathrm{d}\!\log W_{i_w} ,
\end{align}
and boundary values at a point $\vec{s}_0$, 
$\vec{g}^{(w)}_{\tau,\sigma}(\vec{s}_0)$.
The path $\gamma$ connects
$\vec{s}_0$ and $\vec{s}$, and the iteration starts with $[]_{\vec{s}_0} =1$.
The number of integrations $w$ is called the transcendental weight. 
Up to two loops, it suffices to restrict our focus to $w\le 4$.
We choose \cite{Chicherin:2021dyp} 
\begin{align} \label{eq:X0} 
\vec{s}_0 = \left(1,3,2,-2,7,-2 \right) \,,
\end{align} 
which is in the physical $s_{45}$ channel: particles 4 and 5 are in the initial
state, and the remaining particles are in the final state. 
The $\vec g^{(0)}_{\tau,\sigma}(\vec{s}_0)$ are rational numbers and 
are determined by the first-entry condition~\cite{Gaiotto:2011dt} up to
an overall normalization. For $w \geq 1$,
we compute the $\vec g^{(w)}_{\tau,\sigma}(\vec{s}_0)$
with 60-digit precision using 
\texttt{AMFlow}~\cite{Liu:2022chg}, interfaced to 
\texttt{FiniteFlow}~\cite{Peraro:2019svx} and \texttt{LiteRed}~\cite{Lee:2012cn,Lee:2013mka}.

The structure of the iterated integrals in \cref{eq:iterated_integral} is very
well understood~\cite{Chen:1977oja}: integrals involving different
letters are linearly independent, and products of integrals are controlled
by a shuffle algebra.
This enables an algorithmic construction of a
minimal set of functions in which to express the $\vec
g^{(w)}_{\tau,\sigma}(\vec{s})$: the one-mass pentagon functions.
To construct this minimal set, we follow the 
procedure 
in refs~\cite{Chicherin:2020oor,Chicherin:2021dyp}.
We begin by considering the $\vec g^{(w)}_{\tau,\sigma}(\vec{s})$ at symbol
level~\cite{Goncharov:2010jf}
and select a minimal set of the $\vec{g}^{(w)}_{\tau,\sigma}(\vec{s})$ from
the symbol-level solutions, starting at weight 1 and proceeding iteratively up
to weight 4.
To this end, we consider the set of coefficients
$\vec{g}^{(w)}_{\tau,\sigma}(\vec{s})$ for all $\tau$ and $\sigma$, as well as
all weight-$w$ products of lower-weight functions, and select a subset of
linearly independent elements,
preferring products of
lower-weight pentagon functions. This minimizes the number of 
\emph{irreducible functions}: functions which cannot be expressed in
terms of products of lower-weight functions. 
In this way, at each weight we construct a set of algebraically independent
$\vec{g}^{(w)}_{\tau,\sigma}(\vec{s})$, which we call the pentagon functions and
denote by $\{f^{(w)}_i\}$.
We find 11 irreducible functions of weight 1, 35 of weight 2, 
217 of weight 3 and 1028 of weight 4. The total number is substantially lower than
2304, the number of independent master integrals.

We now turn to expressing the
$\vec{g}^{(w)}_{\tau,\sigma}(\vec{s})$ in terms of pentagon functions.
The approach that we use bypasses the
determination of the relations between the boundary values
$\vec g^{(w)}_{\tau,\sigma}(\vec{s}_0)$, a bottleneck of previous approaches.
To this end, we make
the following ansatz: we assume that the $\vec{g}^{(w)}_{\tau,\sigma}$ are
graded polynomials in the pentagon functions and two transcendental constants,
$\zeta_2$ and $\zeta_3$, over the field of rational numbers.
Let us consider a weight-2 coefficient 
${g}^{(2)}_{\tau,\sigma}$ as an example. Our ansatz is
\begin{equation}\label{eq:w2ex}
	{g}^{(2)}_{\tau,\sigma}=
	\sum_{i}a^i_{\tau,\sigma}\,f^{(2)}_i+
	\sum_{i,j}a^{i,j}_{\tau,\sigma}f^{(1)}_if^{(1)}_j
	+\tilde{a}_{\tau,\sigma} \, \zeta_2\,.
\end{equation}
We determine the rational numbers $a^i_{\tau,\sigma}$ and $a^{i,j}_{\tau,\sigma}$
from symbol-level manipulations, 
and the $\tilde{a}_{\tau,\sigma}$ by numerically evaluating
the coefficients on both sides of the equation at $\vec{s}_0$.
In this way, we explicitly write all 
$\vec g^{(w)}_{\tau,\sigma}(\vec{s})$ in terms of $\zeta$~values and
a minimal set of pentagon functions, for which we have an 
iterated-integral representation and a boundary condition valid to 
60 digits.

We emphasize that the ansatz of \cref{eq:w2ex} implies non-trivial polynomial
relations among the $\vec g^{(w)}_{\tau,\sigma}(\vec{s}_0)$.
Previous approaches \cite{Chicherin:2020oor,Badger:2021nhg,Chicherin:2021dyp}
to determine these relations required \texttt{PSLQ}~\cite{PSLQ} studies of
high-precision numerical evaluations ($\mathcal{O}(3000)$ digits) that were
obtained from MPL expressions \cite{Canko:2020ylt}. Bypassing the need for
high-precision evaluations, which are not available for non-planar topologies,
is therefore a substantial improvement.

To validate our results, we checked that they agree with those of 
refs.~\cite{Abreu:2020jxa,Chicherin:2021dyp} for the factorizable and PB families.
For HB families, we compared our results with
evaluations from \texttt{DiffExp}~\cite{Hidding:2020ytt} starting at
the boundary values of ref.~\cite{Abreu:2021smk}. 
We find numerical agreement with the results of ref.~\cite{Kardos:2022tpo} 
for the HBmzz family, but were unable to evaluate their results for HBzzz and HBzmz.
For the two DP families, we started from an evaluation of the pentagon
functions, and used \texttt{DiffExp} to verify that the integrals are regular
at all Euclidean spurious singularities.
We further checked selected permutations of the DP integrals against
\texttt{AMFlow} at random phase-space points.

\section{Numerical Evaluation and Analytic Structure}
\label{sec:pf_evaluation} 

The algorithm given above leaves freedom in the definition of
pentagon functions.
We leverage it to build a basis of functions that
can be evaluated in an efficient and stable way, and
is informed by the singularities that are expected in physical quantities.  
Non-planar families bring considerable new challenges, 
highlighted below.

The general approach to the evaluation of pentagon functions follows 
refs.~\cite{Caron-Huot:2014lda,Gehrmann:2018yef,Chicherin:2020oor,Chicherin:2021dyp}.
We focus on the phase-space region corresponding to the $s_{45}$ channel
defined below \cref{eq:X0}, which is sufficient for hadron-collider processes 
(other $2\to 3$ channels are obtained through appropriate 
permutations~\cite{Chicherin:2021dyp}).
We construct the path $\gamma$ in
eq.~\eqref{eq:iterated_integral} so that it never leaves the $s_{45}$ channel,
following the algorithm of ref.~\cite{Chicherin:2021dyp}.
Up to weight 2, we obtain expressions in terms
of logarithms and dilogarithms~\cite{Duhr:2011zq} with no logarithmic branch points 
within the $s_{45}$ channel.
At weights $3$ and $4$ we construct one-fold integral
representations~\cite{Caron-Huot:2014lda} and perform the integration
numerically.
We refer to ref.~\cite{Chicherin:2021dyp} for a thorough discussion, and
highlight here novel cases where the one-fold
integral representation has a singularity at some point on $\gamma$.

We exemplify our approach with weight-three pentagon
functions, but generalization to weight four is straightforward.
The one-fold integral representations of the
pentagon functions are combinations of terms of the~form
\begin{equation}
  I(h, W) = \int_{\gamma} h(t) \, \partial_t \log (W[t]) \, \mathrm{d}t  \,.
\label{eq:oneFoldIntegral}
\end{equation}
For simplicity, we parametrize $\gamma$ in terms of $t$ so that
$\partial_t \log(W[t])$ diverges at $t=0$ (if it does diverge on $\gamma$).
To construct a numerically stable algorithm for evaluating the pentagon
functions, we consider the analytic structure of \cref{eq:oneFoldIntegral} in
detail. 
In many instances the singularity at $t=0$ cancels in the sum 
of the contributions of the form of \cref{eq:oneFoldIntegral}.
We arrange these cancellations analytically as in refs.~\cite{Chicherin:2020oor,Chicherin:2021dyp}, 
and such pentagon functions are infinitely differentiable in the physical region. 
The novel behavior in this work is a feature of five-point one-mass non-planar pentagon
functions, related to $\Sigma_5^{(i)} = 0$ surfaces.
All cases are summarized in \cref{tab:IntegrandStructure}, and can be organized
in terms of the local behavior of $h$ and $W$.
\begin{table}[tbp]
  \renewcommand{\arraystretch}{1.8}
  \begin{tabular}{C{4ex} *{2}{C{15ex}}*{2}{C{11ex}}}
    \toprule
    case & $\partial_t \log(W[t])$ & $h(t)$ & continuous \\
    \colrule
    $a$\,\, & $\frac{\omega_{1}}{t} + \mathcal{O}(t^{0})$ & $h_{1/2} \sqrt{t} + \mathcal{O}(t)$ & \cmark \\
    $b$\,\, & $\frac{\omega_{1/2}}{\sqrt{t}} + \mathcal{O}(t^{0})$ & $h_0 + \mathcal{O}(\sqrt{t})$ & \cmark \\
    $c$\,\, & $\frac{\omega_{1}}{t} + \mathcal{O}(t^{0})$ & $h_0 + \mathcal{O}(\sqrt{t})$& \xmark  \\
    \botrule
  \end{tabular}
  \caption{Non-differentiable integral functions due to divergent integrands in \cref{eq:oneFoldIntegral}
    at singularity of $\partial_t \log(W[t])$, which cause discontinuities (right column).
  }
  \label{tab:IntegrandStructure}
\end{table}
In all cases the integrands diverge.
For cases $a$ and $b$, the singularity is integrable.
In case $a$, $W = \Sigma_5^{(i \ne 3)}$; in case $b$, $\mathrm{d} \log W$ is odd
with respect to $\sqrt{\Sigma_5^{(3)}}$ and, as $h_0$ is known, 
we handle the integrable singularity with an analytic subtraction.
In case $c$, $W = \Sigma_5^{(3)}$ and the integral has a 
logarithmic singularity.
We introduce a subtraction term and analytically integrate
the $1/t$ singularity, 
resulting in a numerical integration over the remaining integrable
singularity.
We discuss the analytic continuation across $\Sigma_5^{(3)}=0$ in \cref{sec:sigma5-continuation}.
The distinguished role played by $\Sigma_5^{(3)}$ is a consequence of working
in the $s_{45}$ channel.

Integrable singularities in \cref{eq:oneFoldIntegral}
complicate the application of the numerical algorithms of
refs.~\cite{Chicherin:2020oor,Chicherin:2021dyp}. Since
all problematic cases have $W =\Sigma_5^{(i)}$,
we dub the subset of pentagon functions with this behavior as $\fsigma$.
Integrating over the integrable singularities in cases $a$ and $c$ demands higher
intermediate precision, at the cost of performance. This
motivates the construction of the pentagon functions
so that the set $\fsigma$ is as small as possible, which also
isolates the logarithmic singularity at $\Sigma_5^{(3)}=0$ (case $c$) in
as few functions as possible. More generally, we can organize the construction
of the pentagon functions to make analytic properties manifest. For instance, a
conjecture about the absence of the letter ${\Delta_5}$ in the properly
defined finite remainders of scattering amplitudes has been put forward for the massless case
(see~\cite{Chicherin:2020umh,He:2022tph,Bossinger:2022eiy} for possible explanations), 
and we thus isolate
the $\Delta_5$ dependence in as few functions as possible.

\begin{figure}[t]
  \centering
  \begin{subfigure}[t]{0.48\textwidth}
    \centering
    \includegraphics[width=1\linewidth]{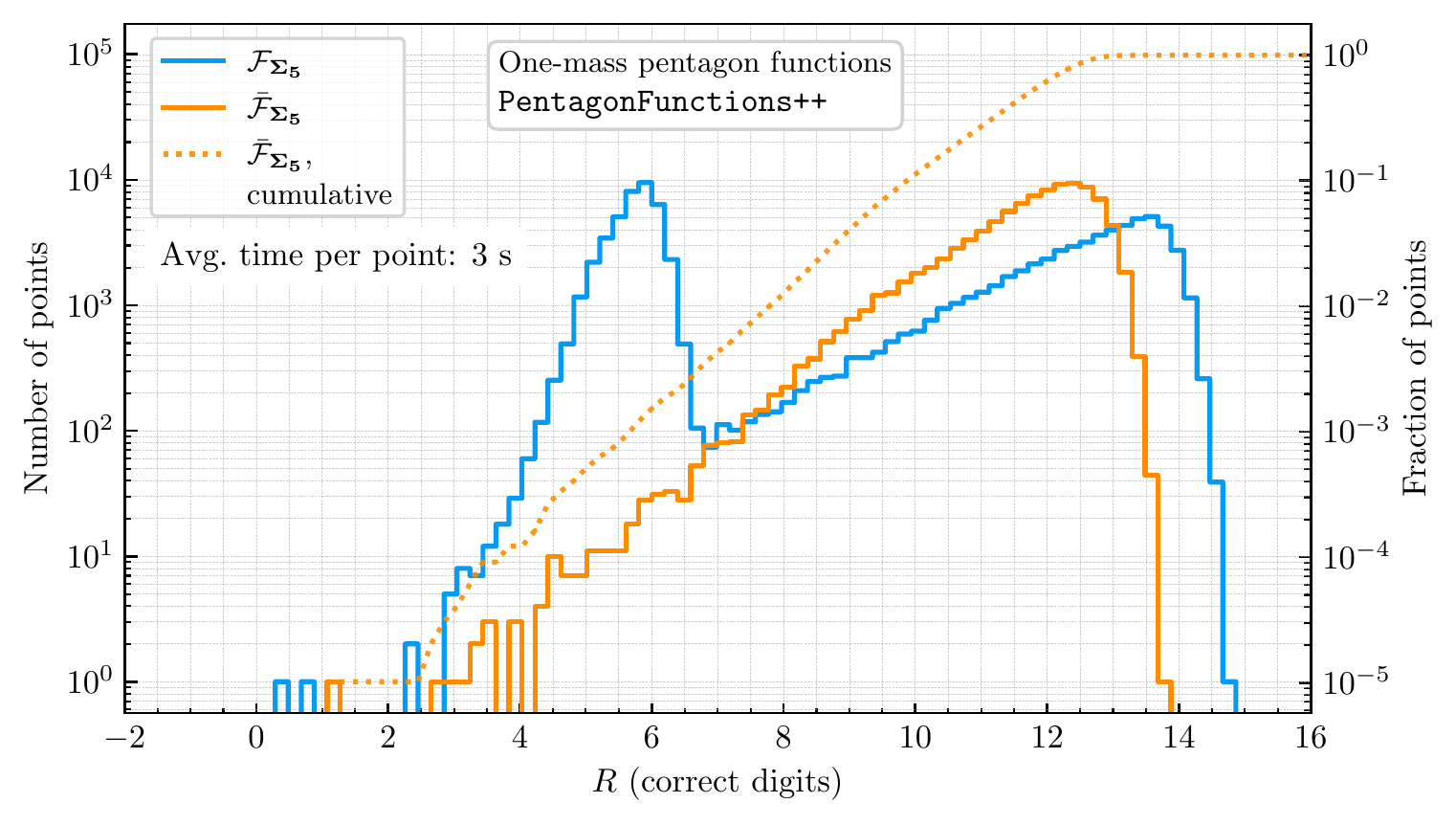}
    \caption{
      Functions $\fsigma$ shown in blue and $\fnosigma$ in orange. 
      Dashed line represents the latter's cumulative distribution.
    } 
    \label{fig:pf_num_stab_sigma5_split}
  \end{subfigure}
  \begin{subfigure}[t]{0.48\textwidth} 
    \centering
    \includegraphics[width=1\linewidth]{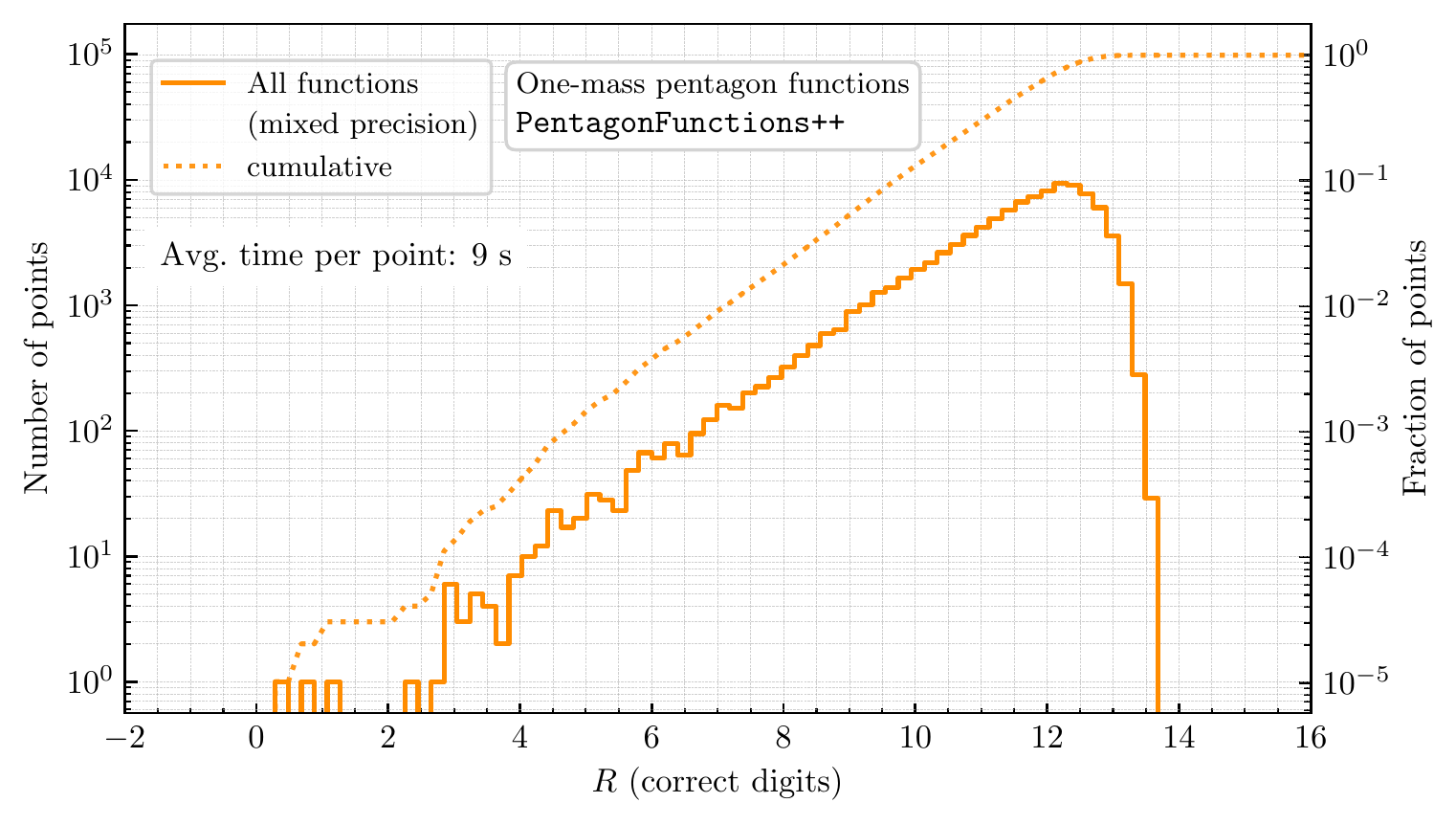}
    \caption{
      All one-mass pentagon functions. 
      The functions $\fsigma$ are evaluated in higher precision when the numerical integration path
      crosses the $\Sigma_5=0$ surface.
    }
    \label{fig:pf_num_stab_full_rescue}
  \end{subfigure}
  \caption{
    Distribution of correct digits compared to quadruple-precision evaluations 
    over 100k points.
  } 
  \label{fig:pf_num_stab}
\end{figure}

The numerical evaluation of the one-mass pentagon
functions is available through the \texttt{C++} library
\texttt{PentagonFunctions++}~\cite{PentagonFunctions:cpp}. To assess
its numerical performance, we evaluate all
functions on a sample of $10^5$ points as in
ref.~\cite{Chicherin:2021dyp}. We present the distributions of correct
digits with standard intermediate precision for the functions in $\fsigma$
(blue line in \cref{fig:pf_num_stab_sigma5_split}) and the ones that
do not involve the letters $\Sigma_5^{(i)}$ (orange line in 
\cref{fig:pf_num_stab_sigma5_split}), denoted $\fnosigma$. 
The blue peak at 6 digits in \cref{fig:pf_num_stab_sigma5_split} is
generated by phase-space points that use integration paths which intersect 
the surfaces $\Sigma_5^{(i)}=0$.
As expected, it is absent for the orange line as well as in
\cref{fig:pf_num_stab_full_rescue}, where higher intermediate precision is
employed to rescue the cases contributing to the blue peak
mentioned above.
Given the evaluation times of a few seconds per point and the overall good 
numerical stability, our results are suitable for immediate phenomenological
applications.

\section{Discussion and Outlook}\label{sec:discussion} 
%

We complete the calculation of all two-loop five-point integrals with massless
propagators and a single massive external leg. Our results allow us to study their
analytic structure, and to 
efficiently evaluate them numerically through weight 4, as needed for current phenomenological
applications. The numerical evaluation is readily available as a
\texttt{C++} library~\cite{PentagonFunctions:cpp}. 

The new algorithm we present for constructing pentagon functions
provides a substantial improvement over 
previous ones, only relying on the knowledge of pure bases and the evaluation of the 
functions at a single point to moderate precision~\cite{Liu:2022chg,Hidding:2022ycg}.
This robust algorithm will certainly
find applications in other classes of integrals.

Our results open the door to further explorations of the analytic structure of this
class of integrals, for instance along the direction of 
refs.~\cite{Hannesdottir:2021kpd,Hannesdottir:2022bmo,Bourjaily:2022vti}.
An undoubtedly important question to explore in more detail is the presence of 
logarithmic singularities within the physical region associated with the letter
$\Sigma_5^{(i)}$ for some master integrals.
Furthermore, our results can be used in the very active area of bootstrapping in 
$\mathcal{N}=4$ sYM (see e.g.\ refs.\ \cite{Dixon:2021tdw,Dixon:2022xqh,Guo:2021bym,Guo:2022qgv,Gopalka:2023doh}).
Finally, they will be central to the calculation of the NNLO corrections to processes
such as the production of a massive boson in association with two jets at hadron 
colliders, as well as the ongoing N$^3$LO calculations \cite{Canko:2021xmn,Henn:2023vbd} 
of processes involving a massive external particle and three massless ones.

\begin{acknowledgements}

D.C., V.S., and S.Z.\ gratefully acknowledge the computing resources provided by 
the Max Planck Institute for Physics. 
We thank Nikolaos Syrrakos for communications 
regarding the results of ref.~\cite{Kardos:2022tpo}, and Andrew McLeod and
Sebastian Mizera for discussions related to the singularities at $\Sigma_5=0$.
This project received funding
from the European Union's Horizon 2020 research and innovation programme 
\textit{High precision multi-jet dynamics at the LHC} (grant agreement No.~772099).
V.S.\ has received funding from the European Research Council (ERC) under the European 
Union's Horizon 2020 research and innovation programme grant agreement 
101019620 (ERC Advanced Grant TOPUP).
The work of B.P.\ was supported by the European Union's Horizon 2020 research and
innovation program under the Marie Sklodowska-Curie grant agreement No.~896690 (LoopAnsatz). D.C.\ is supported by the French National Research Agency in the framework of the \textit{Investissements d’avenir} program (ANR-15-IDEX-02).

\end{acknowledgements}

\appendix

\section{Analytic Continuation Across $\Sigma_5^{(3)}=0$ Surface}
\label{sec:sigma5-continuation}

The analytic continuation of Feynman integrals across singularities
that do not correspond to normal thresholds is known to be a subtle issue.
In this appendix, we discuss how it was addressed in constructing the pentagon
functions defined in this work. More precisely, we focus on the $\Sigma_5^{(3)}=0$
surface which, as noted in the main text, introduces
a logarithmic singularity in the physical region corresponding to the
$s_{45}$ channel. Other similar polynomials, such as $\Delta^{(i)}_3$ and 
$\Delta_5$, have a well defined sign within this region and so do not lead
to singularities~\cite{Chicherin:2021dyp}.

Let us start by commenting on a problem that is related to all square
roots that appear in our alphabet, that is with both $\Sigma_5^{(i)}$ 
as well as $\Delta^{(i)}_3$ and $\Delta_5$.
There are several letters involving the square root of these polynomials. 
These square roots are introduced as normalizations of otherwise rational integrands
and, in practice, we must choose a branch for them. This choice is 
spurious by construction, and should be made consistently in the 
definition of the pentagon functions and in the definition of the pure
basis of integrals. Therefore, without loss of generality, we choose  
the standard principal square root branch, i.e.\ square roots are either 
real positive  or have a positive imaginary part. 
This choice dictates how we evaluate the logarithms 
of algebraic letters in the one-fold integral representations.  
We verified that these logarithms are continuous along the integration path
and cannot pick up any monodromy contributions.
As expected, the dependence on this prescription cancels out in all
quantities which are even under square-root sign changes,
such as scattering amplitudes. 

Let us now return to the main issue that we wish to clarify in this appendix,
namely how to handle integrals that diverge on $\Sigma_5^{(3)}=0$ surfaces
(case $c$ of \cref{tab:IntegrandStructure}).
We start from the iterated integral representation of the pentagon functions 
and manipulate them so that all divergent terms are explicitly written
in terms of
$\log{\Sigma_5^{(3)}}$~\cite{longPaper}.
We must then have a prescription
to analytically continue such terms through the $\Sigma_5^{(3)}=0$ surface.
We define the logarithm as
\begin{equation}\label{eq:contSigma3}
  \log{\Sigma_5^{(3)}} =  \log{\abs{\Sigma_5^{(3)}}} 
  ~+~ \ii \, \pi\,\Theta\left(-\Sigma_5^{(3)} \right) \,,
\end{equation}
where $\Theta$ is the Heaviside step function,
which is consistent with our definition of square roots discussed above.
We proved that this prescription is uniquely fixed by assigning small 
positive imaginary parts to all Mandelstam invariants. 
More precisely, we verified that $\Sigma_5^{(3)}$ receives a small positive imaginary
for an arbitrary path approaching the $\Sigma_5^{(3)}=0$ surface
within the $s_{45}$ channel.
The non-trivial part of the proof involves showing that
all derivatives of $\Sigma_5^{(3)}$ evaluated at $\Sigma_5^{(3)}=0$
are positive in the $s_{45}$ channel. 
The details of this proof are rather lengthy so we do not reproduce them here.

\begin{figure}[ht]
  \centering
  \includegraphics[width=0.7\linewidth]{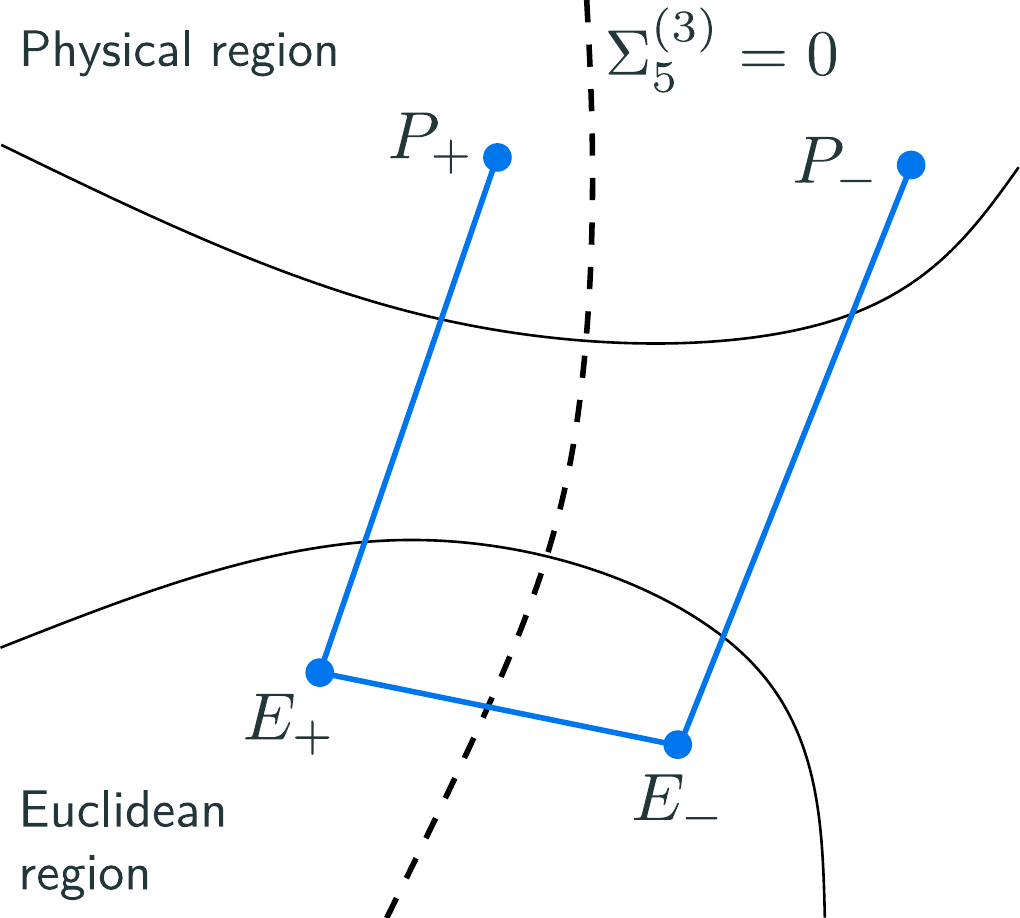}
  \caption{A path that connects phase-space points with opposite $\Sigma_5^{(3)}$ signs without encountering the singularity at the $\Sigma_5^{(3)}=0$ surface.} 
  \label{fig:Sigma5_path}
\end{figure}

\Cref{eq:contSigma3} completely determines
the analytic continuation of the integrals and the pentagon functions
through the $\Sigma_5^{(3)}=0$ surface.
It is, however, instructive to present further evidence for the correctness of this procedure.
We focus on the simplest integral family that contains integrals which are singular at $\Sigma_5^{(3)}=0$, namely HBzmz in the permutation $(3 \to 4,4 \to 5,5 \to 3)$ of the external legs (see fig.~\ref{fig_families_int}).
First, we use \texttt{AMFlow} to evaluate the master integrals at several points on a path approaching the $\Sigma_5^{(3)}=0$ surface within the $s_{45}$ channel.
We observe numerically the onset of the divergence, and that the evaluation of the pentagon functions is in agreement and stable also near the singularity. 
We further evaluate the master integrals at points on the other side of the $\Sigma_5^{(3)}=0$ surface with respect to the base point $\vec{s}_0$ in \cref{eq:X0}, confirming the validity of our prescription in \cref{eq:contSigma3}. 
Second, we use \texttt{DiffExp}~\cite{Hidding:2020ytt} to solve the differential equations along a specially constructed set of paths, 
which are illustrated by \cref{fig:Sigma5_path}.
Starting from $P_+$, where $\Sigma_5^{(3)}>0$, we evolve the master integrals along a path segment $P_+ E_+$ into the Euclidean region without
crossing the $\Sigma_5^{(3)}=0$ surface. We then cross the surface along a path segment $E_+ E_-$, while staying within the Euclidean region where
$\Sigma_5^{(3)}=0$ does not lead to singularities, which guarantees that no analytic continuation through this surface is required. 
Finally, we return to the physical region via a path segment $E_- P_-$, 
along which $\Sigma_5^{(3)}<0$. As noted in \cite{Abreu:2021smk}, this allows one to circumvent
the singular surface, showing that the result of continuing across it is uniquely defined.
We then confirmed that the evaluations of the pentagon functions at the points $P_+$ and $P_-$ agree with the results from \texttt{DiffExp}.

\bibliography{refs} 

\begin{thebibliography}{79}
\expandafter\ifx\csname natexlab\endcsname\relax\def\natexlab#1{#1}\fi
\expandafter\ifx\csname bibnamefont\endcsname\relax
  \def\bibnamefont#1{#1}\fi
\expandafter\ifx\csname bibfnamefont\endcsname\relax
  \def\bibfnamefont#1{#1}\fi
\expandafter\ifx\csname citenamefont\endcsname\relax
  \def\citenamefont#1{#1}\fi
\expandafter\ifx\csname url\endcsname\relax
  \def\url#1{\texttt{#1}}\fi
\expandafter\ifx\csname urlprefix\endcsname\relax\def\urlprefix{URL }\fi
\providecommand{\bibinfo}[2]{#2}
\providecommand{\eprint}[2][]{\url{#2}}

\bibitem[{\citenamefont{Gehrmann et~al.}(2016)\citenamefont{Gehrmann, Henn, and
  Lo~Presti}}]{Gehrmann:2015bfy}
\bibinfo{author}{\bibfnamefont{T.}~\bibnamefont{Gehrmann}},
  \bibinfo{author}{\bibfnamefont{J.~M.} \bibnamefont{Henn}}, \bibnamefont{and}
  \bibinfo{author}{\bibfnamefont{N.~A.} \bibnamefont{Lo~Presti}},
  \bibinfo{journal}{Phys. Rev. Lett.} \textbf{\bibinfo{volume}{116}},
  \bibinfo{pages}{062001} (\bibinfo{year}{2016}), \bibinfo{note}{[Erratum:
  Phys.Rev.Lett. 116, 189903 (2016)]}, \eprint{1511.05409}.

\bibitem[{\citenamefont{Papadopoulos et~al.}(2016)\citenamefont{Papadopoulos,
  Tommasini, and Wever}}]{Papadopoulos:2015jft}
\bibinfo{author}{\bibfnamefont{C.~G.} \bibnamefont{Papadopoulos}},
  \bibinfo{author}{\bibfnamefont{D.}~\bibnamefont{Tommasini}},
  \bibnamefont{and} \bibinfo{author}{\bibfnamefont{C.}~\bibnamefont{Wever}},
  \bibinfo{journal}{JHEP} \textbf{\bibinfo{volume}{04}}, \bibinfo{pages}{078}
  (\bibinfo{year}{2016}), \eprint{1511.09404}.

\bibitem[{\citenamefont{Abreu et~al.}(2019{\natexlab{a}})\citenamefont{Abreu,
  Page, and Zeng}}]{Abreu:2018rcw}
\bibinfo{author}{\bibfnamefont{S.}~\bibnamefont{Abreu}},
  \bibinfo{author}{\bibfnamefont{B.}~\bibnamefont{Page}}, \bibnamefont{and}
  \bibinfo{author}{\bibfnamefont{M.}~\bibnamefont{Zeng}},
  \bibinfo{journal}{JHEP} \textbf{\bibinfo{volume}{01}}, \bibinfo{pages}{006}
  (\bibinfo{year}{2019}{\natexlab{a}}), \eprint{1807.11522}.

\bibitem[{\citenamefont{Abreu et~al.}(2019{\natexlab{b}})\citenamefont{Abreu,
  Dixon, Herrmann, Page, and Zeng}}]{Abreu:2018aqd}
\bibinfo{author}{\bibfnamefont{S.}~\bibnamefont{Abreu}},
  \bibinfo{author}{\bibfnamefont{L.~J.} \bibnamefont{Dixon}},
  \bibinfo{author}{\bibfnamefont{E.}~\bibnamefont{Herrmann}},
  \bibinfo{author}{\bibfnamefont{B.}~\bibnamefont{Page}}, \bibnamefont{and}
  \bibinfo{author}{\bibfnamefont{M.}~\bibnamefont{Zeng}},
  \bibinfo{journal}{Phys. Rev. Lett.} \textbf{\bibinfo{volume}{122}},
  \bibinfo{pages}{121603} (\bibinfo{year}{2019}{\natexlab{b}}),
  \eprint{1812.08941}.

\bibitem[{\citenamefont{Chicherin et~al.}(2019)\citenamefont{Chicherin,
  Gehrmann, Henn, Wasser, Zhang, and Zoia}}]{Chicherin:2018old}
\bibinfo{author}{\bibfnamefont{D.}~\bibnamefont{Chicherin}},
  \bibinfo{author}{\bibfnamefont{T.}~\bibnamefont{Gehrmann}},
  \bibinfo{author}{\bibfnamefont{J.~M.} \bibnamefont{Henn}},
  \bibinfo{author}{\bibfnamefont{P.}~\bibnamefont{Wasser}},
  \bibinfo{author}{\bibfnamefont{Y.}~\bibnamefont{Zhang}}, \bibnamefont{and}
  \bibinfo{author}{\bibfnamefont{S.}~\bibnamefont{Zoia}},
  \bibinfo{journal}{Phys. Rev. Lett.} \textbf{\bibinfo{volume}{123}},
  \bibinfo{pages}{041603} (\bibinfo{year}{2019}), \eprint{1812.11160}.

\bibitem[{\citenamefont{Abreu et~al.}(2020)\citenamefont{Abreu, Ita, Moriello,
  Page, Tschernow, and Zeng}}]{Abreu:2020jxa}
\bibinfo{author}{\bibfnamefont{S.}~\bibnamefont{Abreu}},
  \bibinfo{author}{\bibfnamefont{H.}~\bibnamefont{Ita}},
  \bibinfo{author}{\bibfnamefont{F.}~\bibnamefont{Moriello}},
  \bibinfo{author}{\bibfnamefont{B.}~\bibnamefont{Page}},
  \bibinfo{author}{\bibfnamefont{W.}~\bibnamefont{Tschernow}},
  \bibnamefont{and} \bibinfo{author}{\bibfnamefont{M.}~\bibnamefont{Zeng}},
  \bibinfo{journal}{JHEP} \textbf{\bibinfo{volume}{11}}, \bibinfo{pages}{117}
  (\bibinfo{year}{2020}), \eprint{2005.04195}.

\bibitem[{\citenamefont{Chicherin and Sotnikov}(2020)}]{Chicherin:2020oor}
\bibinfo{author}{\bibfnamefont{D.}~\bibnamefont{Chicherin}} \bibnamefont{and}
  \bibinfo{author}{\bibfnamefont{V.}~\bibnamefont{Sotnikov}},
  \bibinfo{journal}{JHEP} \textbf{\bibinfo{volume}{20}}, \bibinfo{pages}{167}
  (\bibinfo{year}{2020}), \eprint{2009.07803}.

\bibitem[{\citenamefont{Canko et~al.}(2021)\citenamefont{Canko, Papadopoulos,
  and Syrrakos}}]{Canko:2020ylt}
\bibinfo{author}{\bibfnamefont{D.~D.} \bibnamefont{Canko}},
  \bibinfo{author}{\bibfnamefont{C.~G.} \bibnamefont{Papadopoulos}},
  \bibnamefont{and} \bibinfo{author}{\bibfnamefont{N.}~\bibnamefont{Syrrakos}},
  \bibinfo{journal}{JHEP} \textbf{\bibinfo{volume}{01}}, \bibinfo{pages}{199}
  (\bibinfo{year}{2021}), \eprint{2009.13917}.

\bibitem[{\citenamefont{Abreu et~al.}(2022{\natexlab{a}})\citenamefont{Abreu,
  Ita, Page, and Tschernow}}]{Abreu:2021smk}
\bibinfo{author}{\bibfnamefont{S.}~\bibnamefont{Abreu}},
  \bibinfo{author}{\bibfnamefont{H.}~\bibnamefont{Ita}},
  \bibinfo{author}{\bibfnamefont{B.}~\bibnamefont{Page}}, \bibnamefont{and}
  \bibinfo{author}{\bibfnamefont{W.}~\bibnamefont{Tschernow}},
  \bibinfo{journal}{JHEP} \textbf{\bibinfo{volume}{03}}, \bibinfo{pages}{182}
  (\bibinfo{year}{2022}{\natexlab{a}}), \eprint{2107.14180}.

\bibitem[{\citenamefont{Kardos et~al.}(2022)\citenamefont{Kardos, Papadopoulos,
  Smirnov, Syrrakos, and Wever}}]{Kardos:2022tpo}
\bibinfo{author}{\bibfnamefont{A.}~\bibnamefont{Kardos}},
  \bibinfo{author}{\bibfnamefont{C.~G.} \bibnamefont{Papadopoulos}},
  \bibinfo{author}{\bibfnamefont{A.~V.} \bibnamefont{Smirnov}},
  \bibinfo{author}{\bibfnamefont{N.}~\bibnamefont{Syrrakos}}, \bibnamefont{and}
  \bibinfo{author}{\bibfnamefont{C.}~\bibnamefont{Wever}},
  \bibinfo{journal}{JHEP} \textbf{\bibinfo{volume}{05}}, \bibinfo{pages}{033}
  (\bibinfo{year}{2022}), \eprint{2201.07509}.

\bibitem[{\citenamefont{Badger et~al.}(2023)\citenamefont{Badger, Becchetti,
  Chaubey, and Marzucca}}]{Badger:2022hno}
\bibinfo{author}{\bibfnamefont{S.}~\bibnamefont{Badger}},
  \bibinfo{author}{\bibfnamefont{M.}~\bibnamefont{Becchetti}},
  \bibinfo{author}{\bibfnamefont{E.}~\bibnamefont{Chaubey}}, \bibnamefont{and}
  \bibinfo{author}{\bibfnamefont{R.}~\bibnamefont{Marzucca}},
  \bibinfo{journal}{JHEP} \textbf{\bibinfo{volume}{01}}, \bibinfo{pages}{156}
  (\bibinfo{year}{2023}), \eprint{2210.17477}.

\bibitem[{\citenamefont{Hidding and Usovitsch}(2022)}]{Hidding:2022ycg}
\bibinfo{author}{\bibfnamefont{M.}~\bibnamefont{Hidding}} \bibnamefont{and}
  \bibinfo{author}{\bibfnamefont{J.}~\bibnamefont{Usovitsch}}
  (\bibinfo{year}{2022}), \eprint{2206.14790}.

\bibitem[{\citenamefont{Arkani-Hamed et~al.}(2012)\citenamefont{Arkani-Hamed,
  Bourjaily, Cachazo, and Trnka}}]{Arkani-Hamed:2010pyv}
\bibinfo{author}{\bibfnamefont{N.}~\bibnamefont{Arkani-Hamed}},
  \bibinfo{author}{\bibfnamefont{J.~L.} \bibnamefont{Bourjaily}},
  \bibinfo{author}{\bibfnamefont{F.}~\bibnamefont{Cachazo}}, \bibnamefont{and}
  \bibinfo{author}{\bibfnamefont{J.}~\bibnamefont{Trnka}},
  \bibinfo{journal}{JHEP} \textbf{\bibinfo{volume}{06}}, \bibinfo{pages}{125}
  (\bibinfo{year}{2012}), \eprint{1012.6032}.

\bibitem[{\citenamefont{Kotikov}(1991{\natexlab{a}})}]{Kotikov:1990kg}
\bibinfo{author}{\bibfnamefont{A.~V.} \bibnamefont{Kotikov}},
  \bibinfo{journal}{Phys. Lett. B} \textbf{\bibinfo{volume}{254}},
  \bibinfo{pages}{158} (\bibinfo{year}{1991}{\natexlab{a}}).

\bibitem[{\citenamefont{Kotikov}(1991{\natexlab{b}})}]{Kotikov:1991pm}
\bibinfo{author}{\bibfnamefont{A.~V.} \bibnamefont{Kotikov}},
  \bibinfo{journal}{Phys. Lett. B} \textbf{\bibinfo{volume}{267}},
  \bibinfo{pages}{123} (\bibinfo{year}{1991}{\natexlab{b}}),
  \bibinfo{note}{[Erratum: Phys.Lett.B 295, 409--409 (1992)]}.

\bibitem[{\citenamefont{Bern et~al.}(1994)\citenamefont{Bern, Dixon, and
  Kosower}}]{Bern:1993kr}
\bibinfo{author}{\bibfnamefont{Z.}~\bibnamefont{Bern}},
  \bibinfo{author}{\bibfnamefont{L.~J.} \bibnamefont{Dixon}}, \bibnamefont{and}
  \bibinfo{author}{\bibfnamefont{D.~A.} \bibnamefont{Kosower}},
  \bibinfo{journal}{Nucl. Phys. B} \textbf{\bibinfo{volume}{412}},
  \bibinfo{pages}{751} (\bibinfo{year}{1994}), \eprint{hep-ph/9306240}.

\bibitem[{\citenamefont{Remiddi}(1997)}]{Remiddi:1997ny}
\bibinfo{author}{\bibfnamefont{E.}~\bibnamefont{Remiddi}},
  \bibinfo{journal}{Nuovo Cim. A} \textbf{\bibinfo{volume}{110}},
  \bibinfo{pages}{1435} (\bibinfo{year}{1997}), \eprint{hep-th/9711188}.

\bibitem[{\citenamefont{Gehrmann and Remiddi}(2000)}]{Gehrmann:1999as}
\bibinfo{author}{\bibfnamefont{T.}~\bibnamefont{Gehrmann}} \bibnamefont{and}
  \bibinfo{author}{\bibfnamefont{E.}~\bibnamefont{Remiddi}},
  \bibinfo{journal}{Nucl. Phys. B} \textbf{\bibinfo{volume}{580}},
  \bibinfo{pages}{485} (\bibinfo{year}{2000}), \eprint{hep-ph/9912329}.

\bibitem[{\citenamefont{Henn}(2013)}]{Henn:2013pwa}
\bibinfo{author}{\bibfnamefont{J.~M.} \bibnamefont{Henn}},
  \bibinfo{journal}{Phys. Rev. Lett.} \textbf{\bibinfo{volume}{110}},
  \bibinfo{pages}{251601} (\bibinfo{year}{2013}), \eprint{1304.1806}.

\bibitem[{\citenamefont{Dlapa et~al.}(2020)\citenamefont{Dlapa, Henn, and
  Yan}}]{Dlapa:2020cwj}
\bibinfo{author}{\bibfnamefont{C.}~\bibnamefont{Dlapa}},
  \bibinfo{author}{\bibfnamefont{J.}~\bibnamefont{Henn}}, \bibnamefont{and}
  \bibinfo{author}{\bibfnamefont{K.}~\bibnamefont{Yan}},
  \bibinfo{journal}{JHEP} \textbf{\bibinfo{volume}{05}}, \bibinfo{pages}{025}
  (\bibinfo{year}{2020}), \eprint{2002.02340}.

\bibitem[{\citenamefont{Henn et~al.}(2020)\citenamefont{Henn, Mistlberger,
  Smirnov, and Wasser}}]{Henn:2020lye}
\bibinfo{author}{\bibfnamefont{J.}~\bibnamefont{Henn}},
  \bibinfo{author}{\bibfnamefont{B.}~\bibnamefont{Mistlberger}},
  \bibinfo{author}{\bibfnamefont{V.~A.} \bibnamefont{Smirnov}},
  \bibnamefont{and} \bibinfo{author}{\bibfnamefont{P.}~\bibnamefont{Wasser}},
  \bibinfo{journal}{JHEP} \textbf{\bibinfo{volume}{04}}, \bibinfo{pages}{167}
  (\bibinfo{year}{2020}), \eprint{2002.09492}.

\bibitem[{\citenamefont{Dlapa et~al.}(2022)\citenamefont{Dlapa, Henn, and
  Wagner}}]{Dlapa:2022wdu}
\bibinfo{author}{\bibfnamefont{C.}~\bibnamefont{Dlapa}},
  \bibinfo{author}{\bibfnamefont{J.~M.} \bibnamefont{Henn}}, \bibnamefont{and}
  \bibinfo{author}{\bibfnamefont{F.~J.} \bibnamefont{Wagner}}
  (\bibinfo{year}{2022}), \eprint{2211.16357}.

\bibitem[{\citenamefont{G\"orges et~al.}(2023)\citenamefont{G\"orges, Nega,
  Tancredi, and Wagner}}]{Gorges:2023zgv}
\bibinfo{author}{\bibfnamefont{L.}~\bibnamefont{G\"orges}},
  \bibinfo{author}{\bibfnamefont{C.}~\bibnamefont{Nega}},
  \bibinfo{author}{\bibfnamefont{L.}~\bibnamefont{Tancredi}}, \bibnamefont{and}
  \bibinfo{author}{\bibfnamefont{F.~J.} \bibnamefont{Wagner}}
  (\bibinfo{year}{2023}), \eprint{2305.14090}.

\bibitem[{\citenamefont{Dlapa et~al.}(2021)\citenamefont{Dlapa, Li, and
  Zhang}}]{Dlapa:2021qsl}
\bibinfo{author}{\bibfnamefont{C.}~\bibnamefont{Dlapa}},
  \bibinfo{author}{\bibfnamefont{X.}~\bibnamefont{Li}}, \bibnamefont{and}
  \bibinfo{author}{\bibfnamefont{Y.}~\bibnamefont{Zhang}},
  \bibinfo{journal}{JHEP} \textbf{\bibinfo{volume}{07}}, \bibinfo{pages}{227}
  (\bibinfo{year}{2021}), \eprint{2103.04638}.

\bibitem[{\citenamefont{Chen et~al.}(2022)\citenamefont{Chen, Jiang, Ma, Xu,
  and Yang}}]{Chen:2022lzr}
\bibinfo{author}{\bibfnamefont{J.}~\bibnamefont{Chen}},
  \bibinfo{author}{\bibfnamefont{X.}~\bibnamefont{Jiang}},
  \bibinfo{author}{\bibfnamefont{C.}~\bibnamefont{Ma}},
  \bibinfo{author}{\bibfnamefont{X.}~\bibnamefont{Xu}}, \bibnamefont{and}
  \bibinfo{author}{\bibfnamefont{L.~L.} \bibnamefont{Yang}},
  \bibinfo{journal}{JHEP} \textbf{\bibinfo{volume}{07}}, \bibinfo{pages}{066}
  (\bibinfo{year}{2022}), \eprint{2202.08127}.

\bibitem[{\citenamefont{von Manteuffel and
  Schabinger}(2015)}]{vonManteuffel:2014ixa}
\bibinfo{author}{\bibfnamefont{A.}~\bibnamefont{von Manteuffel}}
  \bibnamefont{and} \bibinfo{author}{\bibfnamefont{R.~M.}
  \bibnamefont{Schabinger}}, \bibinfo{journal}{Phys. Lett. B}
  \textbf{\bibinfo{volume}{744}}, \bibinfo{pages}{101} (\bibinfo{year}{2015}),
  \eprint{1406.4513}.

\bibitem[{\citenamefont{Peraro}(2016)}]{Peraro:2016wsq}
\bibinfo{author}{\bibfnamefont{T.}~\bibnamefont{Peraro}},
  \bibinfo{journal}{JHEP} \textbf{\bibinfo{volume}{12}}, \bibinfo{pages}{030}
  (\bibinfo{year}{2016}), \eprint{1608.01902}.

\bibitem[{\citenamefont{Chen}(1977)}]{Chen:1977oja}
\bibinfo{author}{\bibfnamefont{K.-T.} \bibnamefont{Chen}},
  \bibinfo{journal}{Bull. Am. Math. Soc.} \textbf{\bibinfo{volume}{83}},
  \bibinfo{pages}{831} (\bibinfo{year}{1977}).

\bibitem[{Pen()}]{PentagonFunctions:cpp}
\bibinfo{howpublished}{\url{https://gitlab.com/pentagon-functions/PentagonFunctions-cpp}}.

\bibitem[{\citenamefont{Gehrmann et~al.}(2018)\citenamefont{Gehrmann, Henn, and
  Lo~Presti}}]{Gehrmann:2018yef}
\bibinfo{author}{\bibfnamefont{T.}~\bibnamefont{Gehrmann}},
  \bibinfo{author}{\bibfnamefont{J.~M.} \bibnamefont{Henn}}, \bibnamefont{and}
  \bibinfo{author}{\bibfnamefont{N.~A.} \bibnamefont{Lo~Presti}},
  \bibinfo{journal}{JHEP} \textbf{\bibinfo{volume}{10}}, \bibinfo{pages}{103}
  (\bibinfo{year}{2018}), \eprint{1807.09812}.

\bibitem[{\citenamefont{Badger et~al.}(2021{\natexlab{a}})\citenamefont{Badger,
  Hartanto, and Zoia}}]{Badger:2021nhg}
\bibinfo{author}{\bibfnamefont{S.}~\bibnamefont{Badger}},
  \bibinfo{author}{\bibfnamefont{H.~B.} \bibnamefont{Hartanto}},
  \bibnamefont{and} \bibinfo{author}{\bibfnamefont{S.}~\bibnamefont{Zoia}},
  \bibinfo{journal}{Phys. Rev. Lett.} \textbf{\bibinfo{volume}{127}},
  \bibinfo{pages}{012001} (\bibinfo{year}{2021}{\natexlab{a}}),
  \eprint{2102.02516}.

\bibitem[{\citenamefont{Chicherin et~al.}(2022)\citenamefont{Chicherin,
  Sotnikov, and Zoia}}]{Chicherin:2021dyp}
\bibinfo{author}{\bibfnamefont{D.}~\bibnamefont{Chicherin}},
  \bibinfo{author}{\bibfnamefont{V.}~\bibnamefont{Sotnikov}}, \bibnamefont{and}
  \bibinfo{author}{\bibfnamefont{S.}~\bibnamefont{Zoia}},
  \bibinfo{journal}{JHEP} \textbf{\bibinfo{volume}{01}}, \bibinfo{pages}{096}
  (\bibinfo{year}{2022}), \eprint{2110.10111}.

\bibitem[{\citenamefont{Goncharov et~al.}(2010)\citenamefont{Goncharov,
  Spradlin, Vergu, and Volovich}}]{Goncharov:2010jf}
\bibinfo{author}{\bibfnamefont{A.~B.} \bibnamefont{Goncharov}},
  \bibinfo{author}{\bibfnamefont{M.}~\bibnamefont{Spradlin}},
  \bibinfo{author}{\bibfnamefont{C.}~\bibnamefont{Vergu}}, \bibnamefont{and}
  \bibinfo{author}{\bibfnamefont{A.}~\bibnamefont{Volovich}},
  \bibinfo{journal}{Phys. Rev. Lett.} \textbf{\bibinfo{volume}{105}},
  \bibinfo{pages}{151605} (\bibinfo{year}{2010}), \eprint{1006.5703}.

\bibitem[{\citenamefont{Duhr and Brown}(2022)}]{Duhr:2020gdd}
\bibinfo{author}{\bibfnamefont{C.}~\bibnamefont{Duhr}} \bibnamefont{and}
  \bibinfo{author}{\bibfnamefont{F.}~\bibnamefont{Brown}},
  \bibinfo{journal}{PoS} \textbf{\bibinfo{volume}{MA2019}},
  \bibinfo{pages}{005} (\bibinfo{year}{2022}), \eprint{2006.09413}.

\bibitem[{\citenamefont{Steinmann}(1960{\natexlab{a}})}]{Steinmann:thesis}
\bibinfo{author}{\bibfnamefont{O.}~\bibnamefont{Steinmann}},
  \bibinfo{type}{Doctoral thesis}, \bibinfo{school}{ETH Zürich},
  \bibinfo{address}{Zürich} (\bibinfo{year}{1960}{\natexlab{a}}).

\bibitem[{\citenamefont{Steinmann}(1960{\natexlab{b}})}]{Steinmann:1960}
\bibinfo{author}{\bibfnamefont{O.}~\bibnamefont{Steinmann}},
  \bibinfo{journal}{Helvetica Physica Acta} \textbf{\bibinfo{volume}{33}},
  \bibinfo{pages}{347} (\bibinfo{year}{1960}{\natexlab{b}}).

\bibitem[{\citenamefont{Cahill and Stapp}(1975)}]{Cahill:1973qp}
\bibinfo{author}{\bibfnamefont{K.~E.} \bibnamefont{Cahill}} \bibnamefont{and}
  \bibinfo{author}{\bibfnamefont{H.~P.} \bibnamefont{Stapp}},
  \bibinfo{journal}{Annals Phys.} \textbf{\bibinfo{volume}{90}},
  \bibinfo{pages}{438} (\bibinfo{year}{1975}).

\bibitem[{\citenamefont{Caron-Huot et~al.}(2016)\citenamefont{Caron-Huot,
  Dixon, McLeod, and von Hippel}}]{Caron-Huot:2016owq}
\bibinfo{author}{\bibfnamefont{S.}~\bibnamefont{Caron-Huot}},
  \bibinfo{author}{\bibfnamefont{L.~J.} \bibnamefont{Dixon}},
  \bibinfo{author}{\bibfnamefont{A.}~\bibnamefont{McLeod}}, \bibnamefont{and}
  \bibinfo{author}{\bibfnamefont{M.}~\bibnamefont{von Hippel}},
  \bibinfo{journal}{Phys. Rev. Lett.} \textbf{\bibinfo{volume}{117}},
  \bibinfo{pages}{241601} (\bibinfo{year}{2016}), \eprint{1609.00669}.

\bibitem[{\citenamefont{Dixon et~al.}(2017)\citenamefont{Dixon, Drummond,
  Harrington, McLeod, Papathanasiou, and Spradlin}}]{Dixon:2016nkn}
\bibinfo{author}{\bibfnamefont{L.~J.} \bibnamefont{Dixon}},
  \bibinfo{author}{\bibfnamefont{J.}~\bibnamefont{Drummond}},
  \bibinfo{author}{\bibfnamefont{T.}~\bibnamefont{Harrington}},
  \bibinfo{author}{\bibfnamefont{A.~J.} \bibnamefont{McLeod}},
  \bibinfo{author}{\bibfnamefont{G.}~\bibnamefont{Papathanasiou}},
  \bibnamefont{and} \bibinfo{author}{\bibfnamefont{M.}~\bibnamefont{Spradlin}},
  \bibinfo{journal}{JHEP} \textbf{\bibinfo{volume}{02}}, \bibinfo{pages}{137}
  (\bibinfo{year}{2017}), \eprint{1612.08976}.

\bibitem[{\citenamefont{Caron-Huot et~al.}(2018)\citenamefont{Caron-Huot,
  Dixon, von Hippel, McLeod, and Papathanasiou}}]{Caron-Huot:2018dsv}
\bibinfo{author}{\bibfnamefont{S.}~\bibnamefont{Caron-Huot}},
  \bibinfo{author}{\bibfnamefont{L.~J.} \bibnamefont{Dixon}},
  \bibinfo{author}{\bibfnamefont{M.}~\bibnamefont{von Hippel}},
  \bibinfo{author}{\bibfnamefont{A.~J.} \bibnamefont{McLeod}},
  \bibnamefont{and}
  \bibinfo{author}{\bibfnamefont{G.}~\bibnamefont{Papathanasiou}},
  \bibinfo{journal}{JHEP} \textbf{\bibinfo{volume}{07}}, \bibinfo{pages}{170}
  (\bibinfo{year}{2018}), \eprint{1806.01361}.

\bibitem[{\citenamefont{Hannesdottir
  et~al.}(2022{\natexlab{a}})\citenamefont{Hannesdottir, McLeod, Schwartz, and
  Vergu}}]{Hannesdottir:2022xki}
\bibinfo{author}{\bibfnamefont{H.~S.} \bibnamefont{Hannesdottir}},
  \bibinfo{author}{\bibfnamefont{A.~J.} \bibnamefont{McLeod}},
  \bibinfo{author}{\bibfnamefont{M.~D.} \bibnamefont{Schwartz}},
  \bibnamefont{and} \bibinfo{author}{\bibfnamefont{C.}~\bibnamefont{Vergu}}
  (\bibinfo{year}{2022}{\natexlab{a}}), \eprint{2211.07633}.

\bibitem[{\citenamefont{Badger et~al.}(2021{\natexlab{b}})\citenamefont{Badger,
  Hartanto, Kry\'s, and Zoia}}]{Badger:2021ega}
\bibinfo{author}{\bibfnamefont{S.}~\bibnamefont{Badger}},
  \bibinfo{author}{\bibfnamefont{H.~B.} \bibnamefont{Hartanto}},
  \bibinfo{author}{\bibfnamefont{J.}~\bibnamefont{Kry\'s}}, \bibnamefont{and}
  \bibinfo{author}{\bibfnamefont{S.}~\bibnamefont{Zoia}},
  \bibinfo{journal}{JHEP} \textbf{\bibinfo{volume}{11}}, \bibinfo{pages}{012}
  (\bibinfo{year}{2021}{\natexlab{b}}), \eprint{2107.14733}.

\bibitem[{\citenamefont{Abreu et~al.}(2022{\natexlab{b}})\citenamefont{Abreu,
  Febres~Cordero, Ita, Klinkert, Page, and Sotnikov}}]{Abreu:2021asb}
\bibinfo{author}{\bibfnamefont{S.}~\bibnamefont{Abreu}},
  \bibinfo{author}{\bibfnamefont{F.}~\bibnamefont{Febres~Cordero}},
  \bibinfo{author}{\bibfnamefont{H.}~\bibnamefont{Ita}},
  \bibinfo{author}{\bibfnamefont{M.}~\bibnamefont{Klinkert}},
  \bibinfo{author}{\bibfnamefont{B.}~\bibnamefont{Page}}, \bibnamefont{and}
  \bibinfo{author}{\bibfnamefont{V.}~\bibnamefont{Sotnikov}},
  \bibinfo{journal}{JHEP} \textbf{\bibinfo{volume}{04}}, \bibinfo{pages}{042}
  (\bibinfo{year}{2022}{\natexlab{b}}), \eprint{2110.07541}.

\bibitem[{\citenamefont{Badger et~al.}(2022)\citenamefont{Badger, Hartanto,
  Kry\'s, and Zoia}}]{Badger:2022ncb}
\bibinfo{author}{\bibfnamefont{S.}~\bibnamefont{Badger}},
  \bibinfo{author}{\bibfnamefont{H.~B.} \bibnamefont{Hartanto}},
  \bibinfo{author}{\bibfnamefont{J.}~\bibnamefont{Kry\'s}}, \bibnamefont{and}
  \bibinfo{author}{\bibfnamefont{S.}~\bibnamefont{Zoia}},
  \bibinfo{journal}{JHEP} \textbf{\bibinfo{volume}{05}}, \bibinfo{pages}{035}
  (\bibinfo{year}{2022}), \eprint{2201.04075}.

\bibitem[{\citenamefont{Dixon et~al.}(2022)\citenamefont{Dixon, Gurdogan,
  McLeod, and Wilhelm}}]{Dixon:2021tdw}
\bibinfo{author}{\bibfnamefont{L.~J.} \bibnamefont{Dixon}},
  \bibinfo{author}{\bibfnamefont{O.}~\bibnamefont{Gurdogan}},
  \bibinfo{author}{\bibfnamefont{A.~J.} \bibnamefont{McLeod}},
  \bibnamefont{and} \bibinfo{author}{\bibfnamefont{M.}~\bibnamefont{Wilhelm}},
  \bibinfo{journal}{Phys. Rev. Lett.} \textbf{\bibinfo{volume}{128}},
  \bibinfo{pages}{111602} (\bibinfo{year}{2022}), \eprint{2112.06243}.

\bibitem[{\citenamefont{Dixon et~al.}(2023)\citenamefont{Dixon, G\"urdo\u{g}an,
  Liu, McLeod, and Wilhelm}}]{Dixon:2022xqh}
\bibinfo{author}{\bibfnamefont{L.~J.} \bibnamefont{Dixon}},
  \bibinfo{author}{\bibfnamefont{O.}~\bibnamefont{G\"urdo\u{g}an}},
  \bibinfo{author}{\bibfnamefont{Y.-T.} \bibnamefont{Liu}},
  \bibinfo{author}{\bibfnamefont{A.~J.} \bibnamefont{McLeod}},
  \bibnamefont{and} \bibinfo{author}{\bibfnamefont{M.}~\bibnamefont{Wilhelm}},
  \bibinfo{journal}{Phys. Rev. Lett.} \textbf{\bibinfo{volume}{130}},
  \bibinfo{pages}{111601} (\bibinfo{year}{2023}), \eprint{2212.02410}.

\bibitem[{\citenamefont{Hartanto
  et~al.}(2022{\natexlab{a}})\citenamefont{Hartanto, Poncelet, Popescu, and
  Zoia}}]{Hartanto:2022qhh}
\bibinfo{author}{\bibfnamefont{H.~B.} \bibnamefont{Hartanto}},
  \bibinfo{author}{\bibfnamefont{R.}~\bibnamefont{Poncelet}},
  \bibinfo{author}{\bibfnamefont{A.}~\bibnamefont{Popescu}}, \bibnamefont{and}
  \bibinfo{author}{\bibfnamefont{S.}~\bibnamefont{Zoia}},
  \bibinfo{journal}{Phys. Rev. D} \textbf{\bibinfo{volume}{106}},
  \bibinfo{pages}{074016} (\bibinfo{year}{2022}{\natexlab{a}}),
  \eprint{2205.01687}.

\bibitem[{\citenamefont{Hartanto
  et~al.}(2022{\natexlab{b}})\citenamefont{Hartanto, Poncelet, Popescu, and
  Zoia}}]{Hartanto:2022ypo}
\bibinfo{author}{\bibfnamefont{H.~B.} \bibnamefont{Hartanto}},
  \bibinfo{author}{\bibfnamefont{R.}~\bibnamefont{Poncelet}},
  \bibinfo{author}{\bibfnamefont{A.}~\bibnamefont{Popescu}}, \bibnamefont{and}
  \bibinfo{author}{\bibfnamefont{S.}~\bibnamefont{Zoia}}
  (\bibinfo{year}{2022}{\natexlab{b}}), \eprint{2209.03280}.

\bibitem[{\citenamefont{Buonocore et~al.}(2023)\citenamefont{Buonocore, Devoto,
  Kallweit, Mazzitelli, Rottoli, and Savoini}}]{Buonocore:2022pqq}
\bibinfo{author}{\bibfnamefont{L.}~\bibnamefont{Buonocore}},
  \bibinfo{author}{\bibfnamefont{S.}~\bibnamefont{Devoto}},
  \bibinfo{author}{\bibfnamefont{S.}~\bibnamefont{Kallweit}},
  \bibinfo{author}{\bibfnamefont{J.}~\bibnamefont{Mazzitelli}},
  \bibinfo{author}{\bibfnamefont{L.}~\bibnamefont{Rottoli}}, \bibnamefont{and}
  \bibinfo{author}{\bibfnamefont{C.}~\bibnamefont{Savoini}},
  \bibinfo{journal}{Phys. Rev. D} \textbf{\bibinfo{volume}{107}},
  \bibinfo{pages}{074032} (\bibinfo{year}{2023}), \eprint{2212.04954}.

\bibitem[{\citenamefont{Abreu et~al.}(2023)\citenamefont{Abreu, Chicherin, Ita,
  Page, Sotnikov, and Zoia}}]{abreu_samuel_2023_8082812}
\bibinfo{author}{\bibfnamefont{S.}~\bibnamefont{Abreu}},
  \bibinfo{author}{\bibfnamefont{D.}~\bibnamefont{Chicherin}},
  \bibinfo{author}{\bibfnamefont{H.}~\bibnamefont{Ita}},
  \bibinfo{author}{\bibfnamefont{B.}~\bibnamefont{Page}},
  \bibinfo{author}{\bibfnamefont{V.}~\bibnamefont{Sotnikov}}, \bibnamefont{and}
  \bibinfo{author}{\bibfnamefont{S.}~\bibnamefont{Zoia}},
  \emph{\bibinfo{title}{{Supplementary material for ``All Two-Loop Feynman
  Integrals for Five-Point One-Mass Scattering''}}} (\bibinfo{year}{2023}),
  \urlprefix\url{https://doi.org/10.5281/zenodo.8082811}.

\bibitem[{\citenamefont{Abreu et~al.}(2017)\citenamefont{Abreu, Britto, Duhr,
  and Gardi}}]{Abreu:2017ptx}
\bibinfo{author}{\bibfnamefont{S.}~\bibnamefont{Abreu}},
  \bibinfo{author}{\bibfnamefont{R.}~\bibnamefont{Britto}},
  \bibinfo{author}{\bibfnamefont{C.}~\bibnamefont{Duhr}}, \bibnamefont{and}
  \bibinfo{author}{\bibfnamefont{E.}~\bibnamefont{Gardi}},
  \bibinfo{journal}{JHEP} \textbf{\bibinfo{volume}{06}}, \bibinfo{pages}{114}
  (\bibinfo{year}{2017}), \eprint{1702.03163}.

\bibitem[{\citenamefont{Arkani-Hamed and Yuan}(2017)}]{Arkani-Hamed:2017ahv}
\bibinfo{author}{\bibfnamefont{N.}~\bibnamefont{Arkani-Hamed}}
  \bibnamefont{and} \bibinfo{author}{\bibfnamefont{E.~Y.} \bibnamefont{Yuan}}
  (\bibinfo{year}{2017}), \eprint{1712.09991}.

\bibitem[{\citenamefont{Hannesdottir
  et~al.}(2022{\natexlab{b}})\citenamefont{Hannesdottir, McLeod, Schwartz, and
  Vergu}}]{Hannesdottir:2021kpd}
\bibinfo{author}{\bibfnamefont{H.~S.} \bibnamefont{Hannesdottir}},
  \bibinfo{author}{\bibfnamefont{A.~J.} \bibnamefont{McLeod}},
  \bibinfo{author}{\bibfnamefont{M.~D.} \bibnamefont{Schwartz}},
  \bibnamefont{and} \bibinfo{author}{\bibfnamefont{C.}~\bibnamefont{Vergu}},
  \bibinfo{journal}{Phys. Rev. D} \textbf{\bibinfo{volume}{105}},
  \bibinfo{pages}{L061701} (\bibinfo{year}{2022}{\natexlab{b}}),
  \eprint{2109.09744}.

\bibitem[{\citenamefont{Hannesdottir and Mizera}(2022)}]{Hannesdottir:2022bmo}
\bibinfo{author}{\bibfnamefont{H.~S.} \bibnamefont{Hannesdottir}}
  \bibnamefont{and} \bibinfo{author}{\bibfnamefont{S.}~\bibnamefont{Mizera}}
  (\bibinfo{year}{2022}), \eprint{2204.02988}.

\bibitem[{\citenamefont{Bourjaily
  et~al.}(2022{\natexlab{a}})\citenamefont{Bourjaily, Vergu, and von
  Hippel}}]{Bourjaily:2022vti}
\bibinfo{author}{\bibfnamefont{J.~L.} \bibnamefont{Bourjaily}},
  \bibinfo{author}{\bibfnamefont{C.}~\bibnamefont{Vergu}}, \bibnamefont{and}
  \bibinfo{author}{\bibfnamefont{M.}~\bibnamefont{von Hippel}}
  (\bibinfo{year}{2022}{\natexlab{a}}), \eprint{2208.12765}.

\bibitem[{\citenamefont{Tkachov}(1981)}]{Tkachov:1981wb}
\bibinfo{author}{\bibfnamefont{F.~V.} \bibnamefont{Tkachov}},
  \bibinfo{journal}{Phys. Lett. B} \textbf{\bibinfo{volume}{100}},
  \bibinfo{pages}{65} (\bibinfo{year}{1981}).

\bibitem[{\citenamefont{Chetyrkin and Tkachov}(1981)}]{Chetyrkin:1981qh}
\bibinfo{author}{\bibfnamefont{K.~G.} \bibnamefont{Chetyrkin}}
  \bibnamefont{and} \bibinfo{author}{\bibfnamefont{F.~V.}
  \bibnamefont{Tkachov}}, \bibinfo{journal}{Nucl. Phys. B}
  \textbf{\bibinfo{volume}{192}}, \bibinfo{pages}{159} (\bibinfo{year}{1981}).

\bibitem[{\citenamefont{Laporta}(2000)}]{Laporta:2000dsw}
\bibinfo{author}{\bibfnamefont{S.}~\bibnamefont{Laporta}},
  \bibinfo{journal}{Int. J. Mod. Phys. A} \textbf{\bibinfo{volume}{15}},
  \bibinfo{pages}{5087} (\bibinfo{year}{2000}), \eprint{hep-ph/0102033}.

\bibitem[{\citenamefont{Peraro}(2019)}]{Peraro:2019svx}
\bibinfo{author}{\bibfnamefont{T.}~\bibnamefont{Peraro}},
  \bibinfo{journal}{JHEP} \textbf{\bibinfo{volume}{07}}, \bibinfo{pages}{031}
  (\bibinfo{year}{2019}), \eprint{1905.08019}.

\bibitem[{\citenamefont{Lee}(2012)}]{Lee:2012cn}
\bibinfo{author}{\bibfnamefont{R.~N.} \bibnamefont{Lee}}
  (\bibinfo{year}{2012}), \eprint{1212.2685}.

\bibitem[{\citenamefont{Lee}(2014)}]{Lee:2013mka}
\bibinfo{author}{\bibfnamefont{R.~N.} \bibnamefont{Lee}}, \bibinfo{journal}{J.
  Phys. Conf. Ser.} \textbf{\bibinfo{volume}{523}}, \bibinfo{pages}{012059}
  (\bibinfo{year}{2014}), \eprint{1310.1145}.

\bibitem[{\citenamefont{Klappert et~al.}(2021)\citenamefont{Klappert, Lange,
  Maierh\"ofer, and Usovitsch}}]{Klappert:2020nbg}
\bibinfo{author}{\bibfnamefont{J.}~\bibnamefont{Klappert}},
  \bibinfo{author}{\bibfnamefont{F.}~\bibnamefont{Lange}},
  \bibinfo{author}{\bibfnamefont{P.}~\bibnamefont{Maierh\"ofer}},
  \bibnamefont{and}
  \bibinfo{author}{\bibfnamefont{J.}~\bibnamefont{Usovitsch}},
  \bibinfo{journal}{Comput. Phys. Commun.} \textbf{\bibinfo{volume}{266}},
  \bibinfo{pages}{108024} (\bibinfo{year}{2021}), \eprint{2008.06494}.

\bibitem[{\citenamefont{Smirnov and Chuharev}(2020)}]{Smirnov:2019qkx}
\bibinfo{author}{\bibfnamefont{A.~V.} \bibnamefont{Smirnov}} \bibnamefont{and}
  \bibinfo{author}{\bibfnamefont{F.~S.} \bibnamefont{Chuharev}},
  \bibinfo{journal}{Comput. Phys. Commun.} \textbf{\bibinfo{volume}{247 }},
  \bibinfo{pages}{106877} (\bibinfo{year}{2020}), \eprint{1901.07808}.

\bibitem[{\citenamefont{Bourjaily
  et~al.}(2022{\natexlab{b}})\citenamefont{Bourjaily, Langer, and
  Zhang}}]{Bourjaily:2021hcp}
\bibinfo{author}{\bibfnamefont{J.~L.} \bibnamefont{Bourjaily}},
  \bibinfo{author}{\bibfnamefont{C.}~\bibnamefont{Langer}}, \bibnamefont{and}
  \bibinfo{author}{\bibfnamefont{Y.}~\bibnamefont{Zhang}},
  \bibinfo{journal}{JHEP} \textbf{\bibinfo{volume}{08}}, \bibinfo{pages}{176}
  (\bibinfo{year}{2022}{\natexlab{b}}), \eprint{2112.05157}.

\bibitem[{\citenamefont{Gaiotto et~al.}(2011)\citenamefont{Gaiotto, Maldacena,
  Sever, and Vieira}}]{Gaiotto:2011dt}
\bibinfo{author}{\bibfnamefont{D.}~\bibnamefont{Gaiotto}},
  \bibinfo{author}{\bibfnamefont{J.}~\bibnamefont{Maldacena}},
  \bibinfo{author}{\bibfnamefont{A.}~\bibnamefont{Sever}}, \bibnamefont{and}
  \bibinfo{author}{\bibfnamefont{P.}~\bibnamefont{Vieira}},
  \bibinfo{journal}{JHEP} \textbf{\bibinfo{volume}{12}}, \bibinfo{pages}{011}
  (\bibinfo{year}{2011}), \eprint{1102.0062}.

\bibitem[{\citenamefont{Liu and Ma}(2023)}]{Liu:2022chg}
\bibinfo{author}{\bibfnamefont{X.}~\bibnamefont{Liu}} \bibnamefont{and}
  \bibinfo{author}{\bibfnamefont{Y.-Q.} \bibnamefont{Ma}},
  \bibinfo{journal}{Comput. Phys. Commun.} \textbf{\bibinfo{volume}{283}},
  \bibinfo{pages}{108565} (\bibinfo{year}{2023}), \eprint{2201.11669}.

\bibitem[{\citenamefont{Ferguson and Bailey}(1992)}]{PSLQ}
\bibinfo{author}{\bibfnamefont{H.~R.~P.} \bibnamefont{Ferguson}}
  \bibnamefont{and} \bibinfo{author}{\bibfnamefont{D.~H.}
  \bibnamefont{Bailey}}, \bibinfo{journal}{RNR Technical Report RNR-91-032}
  (\bibinfo{year}{1992}).

\bibitem[{\citenamefont{Hidding}(2021)}]{Hidding:2020ytt}
\bibinfo{author}{\bibfnamefont{M.}~\bibnamefont{Hidding}},
  \bibinfo{journal}{Comput. Phys. Commun.} \textbf{\bibinfo{volume}{269}},
  \bibinfo{pages}{108125} (\bibinfo{year}{2021}), \eprint{2006.05510}.

\bibitem[{\citenamefont{Caron-Huot and Henn}(2014)}]{Caron-Huot:2014lda}
\bibinfo{author}{\bibfnamefont{S.}~\bibnamefont{Caron-Huot}} \bibnamefont{and}
  \bibinfo{author}{\bibfnamefont{J.~M.} \bibnamefont{Henn}},
  \bibinfo{journal}{JHEP} \textbf{\bibinfo{volume}{06}}, \bibinfo{pages}{114}
  (\bibinfo{year}{2014}), \eprint{1404.2922}.

\bibitem[{\citenamefont{Duhr et~al.}(2012)\citenamefont{Duhr, Gangl, and
  Rhodes}}]{Duhr:2011zq}
\bibinfo{author}{\bibfnamefont{C.}~\bibnamefont{Duhr}},
  \bibinfo{author}{\bibfnamefont{H.}~\bibnamefont{Gangl}}, \bibnamefont{and}
  \bibinfo{author}{\bibfnamefont{J.~R.} \bibnamefont{Rhodes}},
  \bibinfo{journal}{JHEP} \textbf{\bibinfo{volume}{10}}, \bibinfo{pages}{075}
  (\bibinfo{year}{2012}), \eprint{1110.0458}.

\bibitem[{\citenamefont{Chicherin et~al.}(2021)\citenamefont{Chicherin, Henn,
  and Papathanasiou}}]{Chicherin:2020umh}
\bibinfo{author}{\bibfnamefont{D.}~\bibnamefont{Chicherin}},
  \bibinfo{author}{\bibfnamefont{J.~M.} \bibnamefont{Henn}}, \bibnamefont{and}
  \bibinfo{author}{\bibfnamefont{G.}~\bibnamefont{Papathanasiou}},
  \bibinfo{journal}{Phys. Rev. Lett.} \textbf{\bibinfo{volume}{126}},
  \bibinfo{pages}{091603} (\bibinfo{year}{2021}), \eprint{2012.12285}.

\bibitem[{\citenamefont{He et~al.}(2022)\citenamefont{He, Liu, Tang, and
  Yang}}]{He:2022tph}
\bibinfo{author}{\bibfnamefont{S.}~\bibnamefont{He}},
  \bibinfo{author}{\bibfnamefont{J.}~\bibnamefont{Liu}},
  \bibinfo{author}{\bibfnamefont{Y.}~\bibnamefont{Tang}}, \bibnamefont{and}
  \bibinfo{author}{\bibfnamefont{Q.}~\bibnamefont{Yang}}
  (\bibinfo{year}{2022}), \eprint{2207.13482}.

\bibitem[{\citenamefont{Bossinger et~al.}(2022)\citenamefont{Bossinger,
  Drummond, and Glew}}]{Bossinger:2022eiy}
\bibinfo{author}{\bibfnamefont{L.}~\bibnamefont{Bossinger}},
  \bibinfo{author}{\bibfnamefont{J.~M.} \bibnamefont{Drummond}},
  \bibnamefont{and} \bibinfo{author}{\bibfnamefont{R.}~\bibnamefont{Glew}}
  (\bibinfo{year}{2022}), \eprint{2212.08931}.

\bibitem[{\citenamefont{Guo et~al.}(2021)\citenamefont{Guo, Wang, and
  Yang}}]{Guo:2021bym}
\bibinfo{author}{\bibfnamefont{Y.}~\bibnamefont{Guo}},
  \bibinfo{author}{\bibfnamefont{L.}~\bibnamefont{Wang}}, \bibnamefont{and}
  \bibinfo{author}{\bibfnamefont{G.}~\bibnamefont{Yang}},
  \bibinfo{journal}{Phys. Rev. Lett.} \textbf{\bibinfo{volume}{127}},
  \bibinfo{pages}{151602} (\bibinfo{year}{2021}), \eprint{2106.01374}.

\bibitem[{\citenamefont{Guo et~al.}(2022)\citenamefont{Guo, Wang, and
  Yang}}]{Guo:2022qgv}
\bibinfo{author}{\bibfnamefont{Y.}~\bibnamefont{Guo}},
  \bibinfo{author}{\bibfnamefont{L.}~\bibnamefont{Wang}}, \bibnamefont{and}
  \bibinfo{author}{\bibfnamefont{G.}~\bibnamefont{Yang}}
  (\bibinfo{year}{2022}), \eprint{2209.06816}.

\bibitem[{\citenamefont{Gopalka and Herrmann}(2023)}]{Gopalka:2023doh}
\bibinfo{author}{\bibfnamefont{T.}~\bibnamefont{Gopalka}} \bibnamefont{and}
  \bibinfo{author}{\bibfnamefont{E.}~\bibnamefont{Herrmann}}
  (\bibinfo{year}{2023}), \eprint{2306.13640}.

\bibitem[{\citenamefont{Canko and Syrrakos}(2022)}]{Canko:2021xmn}
\bibinfo{author}{\bibfnamefont{D.~D.} \bibnamefont{Canko}} \bibnamefont{and}
  \bibinfo{author}{\bibfnamefont{N.}~\bibnamefont{Syrrakos}},
  \bibinfo{journal}{JHEP} \textbf{\bibinfo{volume}{04}}, \bibinfo{pages}{134}
  (\bibinfo{year}{2022}), \eprint{2112.14275}.

\bibitem[{\citenamefont{Henn et~al.}(2023)\citenamefont{Henn, Lim, and
  Torres~Bobadilla}}]{Henn:2023vbd}
\bibinfo{author}{\bibfnamefont{J.~M.} \bibnamefont{Henn}},
  \bibinfo{author}{\bibfnamefont{J.}~\bibnamefont{Lim}}, \bibnamefont{and}
  \bibinfo{author}{\bibfnamefont{W.~J.} \bibnamefont{Torres~Bobadilla}},
  \bibinfo{journal}{JHEP} \textbf{\bibinfo{volume}{05}}, \bibinfo{pages}{026}
  (\bibinfo{year}{2023}), \eprint{2302.12776}.

\bibitem[{lon()}]{longPaper}
\bibinfo{note}{S.~Abreu, D.~Chicherin, H.~Ita, B.~Page, V.~Sotnikov,
  W.~Tschernow and S.~Zoia, in preparation.}

\end{thebibliography}

\end{document}